\documentclass{vldb}
\usepackage{balance}  
\usepackage{booktabs} 
\usepackage{graphicx}
\usepackage{enumitem}
\usepackage{xcolor}

\usepackage{totcount}
\regtotcounter{section}

\usepackage[hyphens]{url}
\usepackage{longtable}
\usepackage{multirow}
\usepackage[normalem]{ulem}
\usepackage{subcaption}
\usepackage[ruled,linesnumbered,noend,algo2e,algosection]{algorithm2e}
\usepackage{tabto}
\usepackage[anythingbreaks]{breakurl}
\usepackage{tikz}
\usepackage{todo}
\usepackage[nolineno]{lgrind}

\usepackage{pifont}
\newcommand{\cmark}{\ding{51}}%
\newcommand{\xmark}{\ding{55}}%

\usepackage[toc, page]{appendix}
\usepackage{array}

\newcommand{\SWITCH}[1]{\STATE \textbf{switch} (#1)\begin{ALC@g}}
\newcommand{\ENDSWITCH}{\end{ALC@g}}
\newcommand{\CASE}[1]{\STATE \textbf{case} #1\textbf{:} \begin{ALC@g}}
\newcommand{\ENDCASE}{\end{ALC@g}}
\newcommand{\DEFAULT}{\STATE \textbf{default:} \begin{ALC@g}}
\newcommand{\ENDDEFAULT}{\end{ALC@g}}

\definecolor{darkblue}{RGB}{0,0,90}
\definecolor{darkred}{RGB}{148,0,0}
\definecolor{magenta}{RGB}{93,0,93}
\definecolor{purple}{RGB}{90,0,90}
\definecolor{lightblue}{HTML}{DAE8FC}

\newcommand{\selectq}[1]{\textcolor{blue}{\uppercase{#1}}}
\newcommand{\insertq}[1]{\textcolor{red}{\uppercase{#1}}}
\newcommand{\updateq}[1]{\textcolor{green}{\uppercase{#1}}}
\newcommand{\deleteq}[1]{\textcolor{cyan}{\uppercase{#1}}}

\newcommand*\circled[1]{\tikz[baseline=(char.base)]{
            \node[shape=circle,draw,inner sep=1pt, fill=lightblue] (char) {#1};}}

\begin{document}
\title{{\ttlit Verity}: Blockchains to Detect Insider Attacks in DBMS}

\numberofauthors{3}
\author{
\alignauthor
Shubham S. Srivastava\\
       \affaddr{IIT Kanpur, India}\\
\alignauthor
Medha Atre\\
       \affaddr{University of Oxford, UK}\\
\and  
\alignauthor
Shubham Sharma\thanks{Equal contribution.}\\
       \affaddr{IIT Kanpur, India}\\
\alignauthor
Rahul Gupta\footnotemark[1]\\
       \affaddr{IIT Kanpur, India}\\
\alignauthor
Sandeep K. Shukla\\
       \affaddr{IIT Kanpur, India}\\
}

\maketitle

\newcolumntype{L}[1]{>{\raggedright\let\newline\\\arraybackslash\hspace{0pt}}m{#1}}
\newcolumntype{C}[1]{>{\centering\let\newline\\\arraybackslash\hspace{0pt}}m{#1}}
\newcolumntype{R}[1]{>{\raggedleft\let\newline\\\arraybackslash\hspace{0pt}}m{#1}}

\def\UrlFont{\rmfamily}

\newcommand{\Insert}{\ensuremath{\mathsf{INSERT}}}
\newcommand{\Select}{\ensuremath{\mathsf{SELECT}}}
\newcommand{\Update}{\ensuremath{\mathsf{UPDATE}}}
\newcommand{\Delete}{\ensuremath{\mathsf{DELETE}}}
\newcommand{\sdc}{\ensuremath{\mathsf{SDC}}}
\newcommand{\actiononparse}{\ensuremath{\mathsf{ActionOnParse}}}
\newcommand{\parser}{\ensuremath{\mathsf{Parser}}}
\newcommand{\Parse}{\ensuremath{\mathsf{PARSE}}}
\newcommand{\checking}{\ensuremath{\mathsf{Check}}}
\newcommand{\adjust}{\ensuremath{\mathsf{Adjust}}}
\newcommand{\dbexecute}{\ensuremath{\mathsf{DB\_Exec}}}
\newcommand{\verify}{\ensuremath{\mathsf{Verify}}}

\begin{abstract}
Integrity and security of the data in database systems are typically maintained with access
control policies and firewalls. However, \textit{insider attacks} -- where someone with an
intimate knowledge of the system and
administrative privileges tampers with the data -- pose a unique challenge.
Measures like \textit{append only logging} prove to be insufficient because an attacker with 
administrative privileges can alter logs and login records to eliminate the trace of attack,
thus making insider attacks hard to detect.

In this paper, we propose \textit{Verity} -- first of a kind system to the best of our knowledge.
Verity serves as a \textit{dataless} framework by which \textit{any} blockchain network can be used
to store \textit{fixed-length metadata} about tuples from \textit{any} SQL database, without complete migration
of the database. Verity uses a formalism for parsing SQL queries and query results to check the respective tuples'
integrity using blockchains to detect insider attacks. We have implemented our technique using Hyperledger Fabric,
Composer REST API, and SQLite database. Using TPC-H data and SQL queries of varying complexity
and types, our experiments demonstrate that any overhead of integrity checking
remains constant per tuple in a query's results, and scales linearly.

\end{abstract}

\maketitle

\section{Introduction}\label{intro}
Conventional integrity constraints in a relational database system (DBMS) involve ensuring the integrity of tuples 
according to predefined constraints such as foreign key constraints, data types of attribute values, etc.
Another aspect of integrity stems from malicious tampering of the tuples in a DBMS.
Typically this data integrity is ensured with access control policies and firewalls. With access control policies, 
only selected few users of a DBMS are given administrative privileges. Firewalls ensure that
an \textit{outsider} cannot get direct access to the DBMS server.
However, an \textit{insider attack} is where a privileged user, e.g., an administrator, misuses the
privileges to gain access to or tamper with the data. As reported in the recent 2017 
and 2018 surveys \cite{VormetricReport2015,haystax,catechreport,CollinsCommonSense2016}, about
30\% of organizations face insider attacks and a staggering 55--60\% of the attackers are 
privileged users or administrators. Among the assets that are most at risk, ``database systems''
top the list with 50--57\% of the insider attacks on them. 
The surveys conjecture that the topmost reason for insider attacks is ``insufficient data protection strategies". 
The surveys further point out that 90\% of organizations feel vulnerable to insider threats, and 
insider attacks remain the most difficult to detect. Thus, insider attack has become a non-trivial and non-negligible 
issue in protecting the data integrity in a DBMS.

Insider attacks can be passive or active. Passive attacks involve unethical access and use of the data,
while active attacks involve tampering with the data and logs, to alter the results of queries.
A simple yet illuminating example of the second type of attack is tampering with academic grade records, and it 
has been reported multiple times in the recent past \cite{nyschool,iowa2017,maryland2017,georgia2018}. Some countries like India 
have adopted Electronic Voting Machines (EVMs) for elections, avoiding paper ballots. An EVM has an embedded DBMS 
inside it, and each vote serves as a transaction. These EVMs are also vulnerable to insider attacks and tampering 
\cite{evm3,evm2,evm}. The traditional databases have come a long way over the past several decades in efficient, diverse, 
and scalable data storage solutions. SQL query optimization and processing along with the modern hardware has efficiently 
tackled the ``memory-wall''.
However, as noted above, the new age challenges are security and integrity of the data, and detection and 
prevention of insider attacks forms a critical component.

On the other hand, \textit{blockchain} is an emerging technology for decentralized data storage with strong guarantees of immutability and tamper resistance. Blockchains can be considered analogous to \textit{append only logs} in a native DBMS. However, in 
an insider attack, the attacker with administrative privileges can alter logs and login records to remove proof of data tampering. 
One straight-forward solution could be to push all the databases on blockchain frameworks.
Indeed, there have already been efforts in this direction. E.g., BigchainDB 
\cite{bigchaindb18} integrates Tendermint \cite{tendermint18,tendermint14} with MongoDB \cite{mongodb18} (a 
NoSQL database), and provides a high transaction rate. However, it supports only decentralized blockchain based 
data management eco-system and supports only MongoDB's querying interface. LedgerDB \cite{ledgerdb} is another 
blockchain based database, which supports high transaction throughput. However, LedgerDB supports only a single table 
and does not support various SQL features. ChainDB by Bitpay Inc \cite{chaindb} is another such solution, but it focuses 
on bitcoin transactions, than providing a general purpose DBMS solution. Most
blockchain frameworks, as well as blockchain powered DBMSs, do not provide a rich SQL querying interface that is common to 
a modern DBMS. Additionally, there is a growing concern for data privacy in pushing the existing data on the public blockchain networks 
\cite{privacy}.

Intrusion Detection Systems (IDS) are analytical systems that focus on user profiling for suspicious activity detection, 
e.g., sudden large financial transactions, user logins from irregular locations, transactions non-compliant 
with the DBMS policies etc \cite{demids,bertino05,hu04,srivastava06,lee00,hashemi08,barbara02}. An IDS will not 
necessarily detect an insider attack if the attack does not violate its analytical modelling and user-profiling framework rules,
e.g., a DBMS administrator illegitimately modifying or inserting a few tuples in a DBMS may not come under IDS radar if there is no
\textit{perceived} irregularity of the behaviour.

On this background, we propose a solution to detect insider attacks in a DBMS. The primary contribution of this paper
is \textit{Verity}, that acts as a framework facilitating use of \textit{any} blockchain with \textit{any} SQL DBMS (centralized or
distributed). It uses the data immutability of blockchains, along with the rich SQL 
interface of a DBMS, without requiring to migrate entire data, or adopting a new query interface or language.
The main novelty of Verity lies in our detailed algorithmic and protocol framework to handle a variety of CRUD (Create, Read, Update, 
Delete) SQL queries by supporting a large part of the SQL grammar -- specifically the queries having nested \texttt{SELECT}
clauses and joins over multiple tables (Section \ref{sec:arch}, Section \ref{sec:sql}).
Verity layer in itself does not store any data or metadata about data, thus maintaining data privacy.
It facilitates identification of illegitimate data tampering whenever the tampered data 
is accessed in an SQL query evaluation. This in turn helps to stop any \textit{cascade} effect of the tampered data
affecting critical decisions based on it, e.g., academic grades, financial accounting etc\footnote{\scriptsize{Prevention
of an  insider attack will require different artifacts for interacting with the DBMS, and is not within the scope of
the current paper.}}. Our solution also allows the flexibility of enabling or disabling the blockchain-based integrity 
checking in a \textit{plug-and-play} fashion.
We have implemented Verity using web-based SQL interface, Hyperledger Fabric \cite{fabric}, 
its Composer REST API \cite{restapi}, and SQLite database \cite{sqlite}.
Since Verity is a framework facilitating the use of blockchains with an SQL DBMS, for the scope of this paper, we have not 
focused on the aspects of performance optimization for increasing the system throughput, 
failure-recovery, or efficient storage/indexing methods, because they are dependent on individual blockchain and DBMS 
platforms. However, our experimental results on TPC-H data of varying scaling factor and SQL queries of
varying complexity (Section \ref{sec:expt}, Appendix \ref{appendix:sql})
demonstrate that any overhead incurred by Verity's integrity checking process remains constant per tuple in the results of a query, 
and thus scales linearly. Also in Section \ref{sec:discussion} we discuss the aspects of possible future throughput optimizations.

\section{Problem Setting} \label{sec:setting}

In this section, we discuss preliminaries about blockchain framework, cryptographic hash functions, and data privacy,
along with an overview of the attacker model and proposed framework.

\subsection{Blockchains} \label{sec:blockchains}
The concept of blockchains was introduced as a technology powering a
peer to peer digital currency, \textit{Bitcoin} 
\cite{nakamoto}.
However, capabilities of blockchains go beyond cryptocurrencies, as it can be used as a decentralized data store with strong
cryptographic guarantees of tamper resistance.
Conceptually, a blockchain is a linked-list, where each node in the list is called a \textit{block}.
Each block is \textit{cryptographically linked} to the previous block, 
forming a continuously growing \textit{chain of blocks}. The first block is called {\it Genesis Block}, which is 
known to all the blockchain participants, and acts as a common reference for verifying the sanctity of the blockchain. Each block after genesis 
may contain several {\it transactions} and other \textit{metadata} such as timestamp, \textit{block height} 
(distance from the Genesis Block) etc. 
Unlike a conventional data storage or DBMS, blockchain has stronger cryptographic guarantees, i.e., data once written in a block,  cannot be easily modified, and creation of a new block on the chain requires
consensus among blockchain peers.
Any roll-back of a transaction or erasure of data gets logged as another transaction in a block, thus providing strong guarantees of \textit{data provenance} (traceability).

In a blockchain, all the participants are \textit{peers}. 
Identities of these peers may be known or hidden, which makes two broad categories of blockchains: (a) \textit{permissionless} with hidden peer identities, and (b) \textit{permissioned} with known peer identities. Permissionless blockchains are more popular for cryptocurrencies, where any peer can join the blockchain network and participate in the consensus protocol. Permissioned blockchains are more suitable for business applications, and they provide an alternate way of peer 
consensus than permissionless blockchains.
Among permissioned blockchains, each may have different type of peer consensus. In Verity, we have used Hyperledger Fabric \cite{fabric}. However, Verity by design is agnostic to any specific blockchain,
and if a different blockchain network is used, 
appropriate peer consensus protocol will need to be implemented.
For the scope of this paper, we do not get into the details of peer consensus algorithms, and refer the interested reader to the relevant literature \cite{nakamoto,fabric}.

However, we would like to make  a note here that unlike a conventional DBMS, where even a single administrator or a small number of them colluding can tamper the data and logs, in a blockchain, it requires collusion among a large fraction of peers, making it a non-trivial process. We explain this further in Section \ref{sec:attackmodel}.

\subsection{Attacker Model Assumptions} \label{sec:attackmodel}
A DBMS has one or more administrators, and we assume that an inside attacker has full access to the database server with 
administrative privileges. Thus the administrator/s can tamper with the data, and modify logs and login records as well.
A typical DBMS and its administrators do not have a strong cryptographic \textit{peer consensus} protocol like blockchains
to authorize a change. Any subset of the administrators can pose a threat as inside attackers.

The goal of Verity is to detect an insider attack, so we assume that the set of peers in a permissioned blockchain that we use along with a DBMS, are not exactly same as the respective database administrators. We also assume that, in a 
rare case, if the permissioned blockchain peers are same as DBMS administrators, \textit{not all} of them are colluding in an insider 
attack. An insider attack that may go undetected for a long time can jeopardize any decisions based on the tampered data due to the 
\textit{cascading} effect. In this scenario, we also assume that in an organization, not all critical decision makers are DBMS 
administrators -- for the sake of fairness, the three sets (1) blockchain peers, (2) DBMS administrators, and (3) critical decision 
makers are \textit{not} completely overlapping one another -- thus making collusion among a large fraction of them less likely.

Our goal is to make tamper detection automatic, whenever the tampered data is accessed through SQL queries, than relying 
on manual audit, and without involving any specific privileged users. We also want tampering to be 
visible to anyone having access to the concerned data. In Section \ref{sec:arch} we demonstrate how we 
achieve this.

For the scope of Verity, we assume that administrators' access methods, such as private keys or passwords are secure.
Our attack model does not consider any application vulnerability or system vulnerability based attacks --
we consider that otherwise the database is \textit{not hacked} -- only a
privileged user is tampering with the tuples. Note, however, that since Verity uses blockchains to log changes and verify
tuple integrity, as described further in this paper, even data tampering done via hacking and bypassing blockchains
can be detected. Peer consensus makes logging an illegitimate change on blockchain non-trivial, and
adds traceability to it. Any aggressive, active damage to the system such as corrupting the hardware or physical 
theft of property is self-evident, and will result in detection of the attack.

\subsection{Cryptographic Hash Functions}  \label{sec:crypto}
In the blockchain nodes we mainly store \textit{metadata} about tuples in a DBMS. This process is elaborated in Section \ref{sec:metadata}. In this section, we set the basic background of hash functions used to generate this metadata.

Cryptographic hash functions \cite{chf,Handschuh2005} map any data of arbitrary 
length to a bit string of fixed length, called {\it hash}. 
A hash function is given as: $h: \{0,1\}^* \rightarrow \{0,1\}^n$ where $n$ is a constant such as 128, 256, or 512. 
By design, cryptographic hash functions are \textit{one-way} functions, and it is computationally difficult to 
\textit{invert} (a.k.a. \textit{break}) them, to generate the original data from a given hash.  
Also for two different pieces of input data, the difference in 
their hash values $H_1, H_2$ is not relative, and often a minor change in the input value results in a drastic change 
in the output hash value.
Noting these properties of cryptographic hash functions, in Verity we use SHA256 \cite{Handschuh2005} to generate a unique \textit{fingerprint} of 256-bit for each tuple in the DBMS. 
This procedure is elaborated in the next section.

\subsection{Metadata about Data}\label{sec:metadata}
In order to use a blockchain network with an existing DBMS, without migrating the whole data on the blockchain, we create \textit{metadata about data}, and only store that on the blockchain.
This metadata is generated using the SHA256 cryptographic hash function for each tuple in a DBMS as follows.
We assume that each table has a primary key defined on it\footnote{\small{If a table does not have a primary key, we can use 
row-number as the primary key, or treat all the columns together as a composite primary key.}}.
For each tuple $r$ in table $T$, we first generate \textit{RowID} as: 
$$\mathbf{RowID}_{r,T} =  h(PrimaryKey(r)\cdot T)$$ 
That is we concatenate  primary key of the tuple along with the table name and generate a hash from it.
Assuming that the tuple has in all $k$ columns, $c_1...c_k$, 
using the \textit{RowID} generated earlier, we generate \textit{fingerprint} of the entire tuple as follows:
\[
 \mathbf{fingerprint_r} = h(\mathit{RowID}\cdot c_1 \cdot c_2 \cdot c_3 \cdots c_k)
\]
That is, we concatenate the \textit{RowID} along with each column value and generate another 256-bit hash from it.
If an attribute has a NULL value, we skip over that and concatenate the next non-NULL valued attribute.
This is specifically done so that any change of a NULL value to a non-NULL and vice versa can be detected through the change in 
the fingerprint.

This fingerprint is then stored on a blockchain using digital signature (private key) of
a blockchain \textit{peer}, and after passing the peer consensus protocol (ref. Section \ref{sec:blockchains}). Thus every 
fingerprint of every tuple stored in a blockchain has an owner (blockchain peer) of the transaction associated with it. Any 
legitimate modifications to a tuple get logged on the blockchain with a new fingerprint of the tuple along with the digital signature 
of the respective owner. 
Any previous fingerprints of the modified tuple are also preserved in the blockchain. Thus blockchain provides non-repudiation
on maintaining the number of legitimate tuple modifications along with their owners. 
However, with hash based fingerprints, the exact nature of modifications cannot be tracked through blockchains. This is in order to 
honour data privacy as discussed next.

\subsection{Data Privacy} \label{sec:privacy}
In our setting, the aforementioned fingerprint of each tuple in a DBMS is pushed on the blockchain. Along with this, we also  store number of rows in each table on the blockchain.
Detailed handling of this metadata in the event of updates is discussed in 
Section \ref{sec:arch}. Note that our method of storing fingerprints preserves the privacy of the original data in a DBMS, because from a fingerprint of a tuple, original attribute values cannot be retrieved (ref. Section \ref{sec:crypto}).
We trust that this is an important property of Verity system, especially among the growing concern of data privacy in using public blockchain frameworks \cite{privacy}. Thus with Verity, even if a public blockchain framework is used for storing metadata, the 
original database values are never revealed on the blockchain framework. 

\subsection{Detecting Tampering} \label{sec:detection}
Conceptually, working of Verity can be summarized as follows:
\begin{enumerate}[itemindent=0mm,labelindent=0mm,leftmargin=4mm,noitemsep]
 \item We assume that at the very beginning the data in a DBMS is clear from any tampering or existing attack.
 
 \item We create a fingerprint for each tuple, and store this fingerprint on the 
blockchain.
 
 \item As given in Section \ref{sec:attackmodel}, we assume that the same set of DBMS administrators are
  not peers on the blockchain network, and not all blockchain peers are colluding in an attack.
 
 \item Further we assume that a normal user does not have a way to access/query the given DBMS by circumventing the
Verity framework. Note that in Section \ref{sec:attackmodel}, we have clarified that in the context of Verity, we are \textit{not} 
assuming a \textit{hacker} scenario, and private keys and passwords are secure.
 
 \item A well-intentioned DBMS administrator makes any updates to the data through Verity, following the peer consensus 
protocol, and the modification gets logged into the blockchain. Each update that gets logged 
on the blockchain has a digital signature of the peer submitting the transaction.
Thus for every blockchain update, there is accountability, and number of modifications can be traced through blockchain.
 
 \item When an inside attacker tampers with the DBMS by circumventing Verity and in turn blockchain logging, the 
tampering gets detected when the tampered tuples are retrieved in an SQL query at a later time issued through Verity's interface.
\end{enumerate}

Having summarized the overall concept and functioning of Verity, next in Section \ref{sec:arch}, we elaborate on 
the detailed architecture, and handling of SQL queries by taking into consideration SQL grammar.
We trust that this will facilitate the community to use the powerful features of any commercial or opensource blockchain framework with any SQL database, without having to migrate entirely to a new system. This, in our opinion, will greatly help in faster detection of insider attacks.

\section{Architecture} \label{sec:arch}

Three main components of the Verity framework are:
\begin{enumerate}[label=(\alph*),leftmargin=8mm,noitemsep]
 \item A blockchain network,
 \item An SQL database, and
 \item An HTTP based web application connecting these two that intercepts SQL queries for 
data integrity checks.
\end{enumerate}

\begin{figure}[t]
\centering
\includegraphics[scale=0.45]{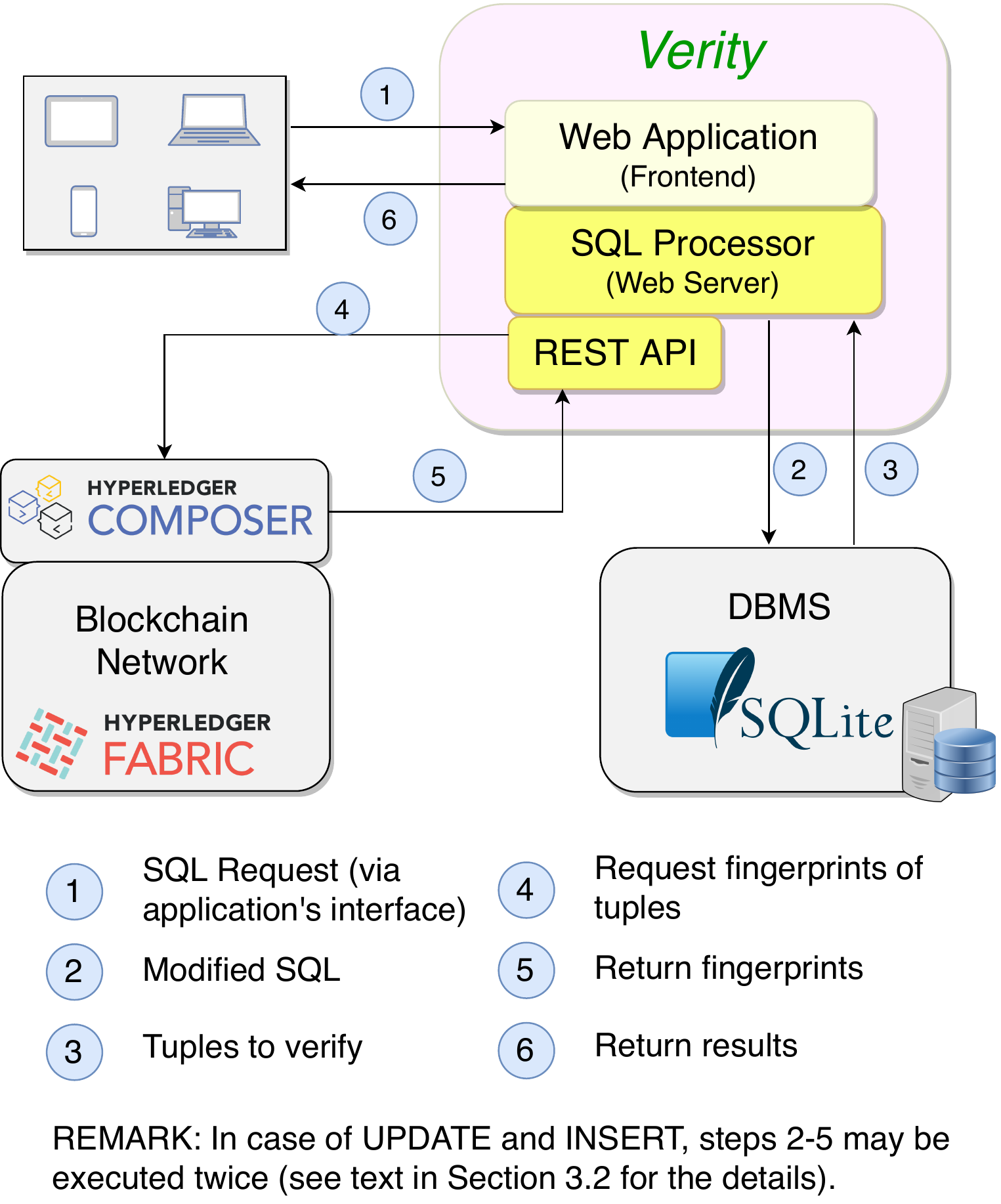}
\caption{Architecture and Communication Sequence}\label{fig:archprot}
\end{figure}

Blockchain network and SQL database can be any blockchain and any DBMS.
It is the third component (c) where we make our main contributions. Through this component we facilitate a DBMS to use 
a blockchain network \textit{without} requiring to migrate entire data on the blockchain. Verity's system
architecture is given in Figure \ref{fig:archprot}.
The web application server is an HTTP based frontend for users to issue SQL queries.

In this interface, we have built our logic of parsing and intercepting SQL queries issued by an end user, and 
checking fingerprints of the tuples involved in building the  results of that query against blockchain. REST API is an interface for Hyperledger Composer framework and is used for querying Hyperledger Fabric for retrieving or adding tuple fingerprints (ref. Section \ref{sec:metadata}).

\textbf{Note:} In Verity, we have used an HTTP based interface for SQL query processing, but this interface can be in any other form too. E.g., command line or programmatic (e.g., JDBC, ODBC), based on the individual application need. Using the detailed SQL parsing algorithms given further in this section, any different interface can be built for achieving the same functionality.

In Section \ref{sec:protocol} we take a brief overview of the sequence of actions taken in the Verity framework for any SQL query issued through it, and then in Section \ref{sec:sql} we give the details of handling four main type of SQL queries -- SELECT, INSERT, UPDATE, DELETE, and any intermix of them adhering to the SQL grammar.

\subsection{Verity Communication Sequence} \label{sec:protocol}

Figure \ref{fig:archprot} shows the interactions between different components of Verity for any SQL query.
An SQL query initiated using the web application interface (\circled{1}) is parsed by the SQL processor, and it is sent to the DBMS with methodical modifications required to check integrity (\circled{2}). The details of this process are discussed in Section \ref{sec:sql}.
DBMS returns tuples matching this modified query (\circled{3}). These tuples are then verified using the corresponding fingerprints stored on the blockchain (\circled{4}, \circled{5}), which constitutes Verity's {\it integrity checking phase}.
Once the check is successfully completed, Verity returns results of the \textit{original} SQL query to the end user.
Thus the integrity checking process is completely opaque to the end user, and user is notified only if the integrity check fails. In case of INSERT, UPDATE, DELETE queries, steps \circled{2}--\circled{5} may have to be executed twice if these queries have nested SELECT queries inside them.
This process is elaborated further in Section \ref{sec:sql}.

\subsection{SQL Parsing} \label{sec:sql}

As noted before in Section \ref{sec:detection}, our
tamper detection model is through intercepting SQL queries and their results to check tuples' integrity.
In this section we elaborate on this aspect and show methodically how tampering can be detected through
subsequent SQL query results using the Verity framework. We achieve this in the following main steps.

\begin{enumerate}[itemindent=0mm,labelindent=0mm,leftmargin=4mm,noitemsep]
 \item Parse the given SQL query using standard SQL grammar.
 
 \item Determine which tuples from the base tables are accessed, modified, or inserted in the query execution and results.
 For this step, Verity maintains information about the DBMS table schemas with it. Note that consistent with Verity's privacy policy, 
 it only stores DBMS schema, and not actual data or metadata (fingerprints of the tuples).
 
 \item Depending on the type of the query (detailed in Sections \ref{sec:select}--\ref{sec:delete}) retrieve the tuples accessed or modified by the query from the base tables in \textit{entirety}, i.e., with all the attributes of those tuples. This step is opaque to the end user.
 
 \item Generate fingerprints of these tuples as given in Section \ref{sec:metadata}, and cross-check those fingerprints against 
 Hyperledger Fabric, or if tuples are inserted or modified, log the new fingerprints on Hyperledger.
 
 \item Once the validity of all the base tuples is established, send back the results of the \textit{original} user query.
\end{enumerate}

SQL is a rich data manipulation language with a lot of syntactic sugar. Within the entire range of SQL's syntactic features, 
currently in Verity, we have focused on SELECT, INSERT, UPDATE, and DELETE queries, with any intermix of them. Presently we do not 
handle queries with outerjoins, and \texttt{IN}, \texttt{ANY}, \texttt{EXISTS}, \texttt{GROUP-BY}, \texttt{HAVING}, and aggregation clauses. In the 
future we plan to extend our parser to handle most of the SQL syntax.

In Verity, we do not assume any special access to the DBMS for knowing the tuples processed by an SQL query.
Thus it is imperative to methodically intercept and parse every SQL query to get the 
base tuples for the validity check. We have achieved this through a \textit{LookAhead Left-to-Right} (LALR) bottom up 
SQL parser. This is similar to a typical DBMS query parser and
plan generator.

Next we present our algorithms to handle the four type of queries with any intermix and nesting. Algorithm \ref{alg:parser}
is the general outer wrapper which accepts an SQL query and invokes appropriate parsing mechanism depending
on the type of the \textit{outermost} query, i.e., if an INSERT query has a nested SELECT subquery, the parser calls
Algorithm \ref{alg:insert} meant to process INSERT queries (ref line \ref{alg:parser:insert} in Algorithm \ref{alg:parser}).

\begin{algorithm2e}
	\SetAlgoLined
	
	\Switch{type of Q}{
	  \Case(Algorithm \ref{alg:select}){SELECT}{
		\Select(Q, $\phi$)\;
	  }
  	  \Case(Algorithm \ref{alg:update}){UPDATE}{
		\Update(Q)\;
		
	  }
 	  \Case(Algorithm \ref{alg:insert}){INSERT}{
		\Insert(Q)\; \label{alg:parser:insert}
		
	  }
  	  \Case(Algorithm \ref{alg:delete}){DELETE}{
		\Delete(Q)\;
		
	  }
	}
	\caption{Parser(Q)} \label{alg:parser}
\end{algorithm2e}

\subsubsection{SELECT queries} \label{sec:select}
SQL grammar for a SELECT query is given below (complete grammar 
is not shown for the sake of simplicity).
\begin{verbatim}
  select_statement: `SELECT' projection-attr `FROM'
                     tables [`WHERE' qualifications]
  tables: single_table | `(' select_statement `)'
                    | tables `,' tables
\end{verbatim}

An SQL query -- with or without joins -- can have projections for only some attributes from the 
\texttt{tables}, and SQL grammar allows \texttt{FROM} clause to have a nested SELECT query, which is treated as a temporary 
table\footnote{\scriptsize{Since currently we do not handle \texttt{IN}, \texttt{ANY}, \texttt{EXISTS}, 
\texttt{GROUP-BY}, \texttt{HAVING} clauses, we assume that the \texttt{WHERE} clause will be devoid of any nested SELECT 
statements.}}. However, recall from Section \ref{sec:metadata} that the \textit{fingerprint} of each tuple stored on the blockchain 
includes all the attributes in that tuple. Thus to verify integrity of a tuple, we need to retrieve
all the attributes of it from the base tables. One na\"{\i}ve way of doing it is to retrieve all the tuples in the base tables before query 
execution, and check each tuple's integrity. However, this incurs following main problems -- (a) it misses the advantages of 
\textit{selectivity} of a query\footnote{\scriptsize{Selectivity of a query is high if it accesses only few tuples, and vice 
versa.}}, (b) it puts the onus of performing joins of these tuples after verification on the Verity framework, thus missing the 
benefits of SQL query optimization methods of a native DBMS, (c) it incurs the problem of ``atomicity'' of checking the integrity of tuples and returning query results to the user -- commonly known as the \textit{Time Of Check to Time Of Use} (TOCTOU) race condition. Hence we perform this step as follows.

As shown in the grammar, a SELECT query can have nested subqueries within it as a part of the \texttt{FROM} clause . Our LALR parser does \textit{bottom-up} parsing of all the subqueries wherein innermost 
nested SELECT query is parsed first.
For every detected SELECT query, the parser modifies its projected attributes to include all the attributes of the respective base tables in 
the \texttt{FROM} clause. Consider for instance, the following query on tables $t1, t2, t3$ with attributes $(x, y, a)$, 
$(a, s, b)$, $(c, d, b)$ respectively stored in that order.

\vspace{2mm}
\begin{lgrind}
\L{\LB{\K{SELECT}_\V{t1}.\V{a},_\V{t2}.\V{b},_\V{t3}.\V{c}}}
\L{\LB{\K{FROM}_\V{t1},_\V{t2},_\V{t3}}}
\L{\LB{\K{WHERE}_\V{t1}.\V{a}=\V{t2}.\V{a}_\K{AND}_\V{t2}.\V{b}=\V{t3}.\V{b}}}
\end{lgrind}
\vspace{2mm}

This query has two joins over three tables, $t1, t2, t3$, and is projecting out only three out of total nine attributes in three tables. For Verity's integrity check however, we need to have all the attributes of the tuples from $t1, t2, t3$ that are part of the join results. Thus Verity's SQL parser internally rewrites this query as follows.

\vspace{2mm}
\begin{lgrind}
\L{\LB{\K{SELECT}_\V{t1}.\V{x},_\V{t1}.\V{y},_\V{t1}.\V{a},_\V{t2}.\V{a},_\V{t2}.\V{s},}}
\L{\LB{}\Tab{11}{\V{t2}.\V{b},_\V{t3}.\V{c},_\V{t3}.\V{d},_\V{t3}.\V{b}}}
\L{\LB{\K{FROM}_\V{t1},_\V{t2},_\V{t3}}}
\L{\LB{\K{WHERE}_\V{t1}.\V{a}=\V{t2}.\V{a}_\K{AND}_\V{t2}.\V{b}=\V{t3}.\V{b}}}
\end{lgrind}
\vspace{2mm}

Now consider the same query rewritten in a different syntax using  a nested SELECT clause as given below.

\vspace{2mm}
\begin{lgrind}
\L{\LB{\K{SELECT}_\V{t1}.\V{a},_\V{r1}.\V{b},_\V{t3}.\V{c}}}
\L{\LB{\K{FROM}_\V{t1},_(\K{SELECT}_\V{a},_\V{b}_\K{FROM}_\V{t2})_\K{AS}_\V{r1},_\V{t3}}}
\L{\LB{\K{WHERE}_\V{t1}.\V{a}=\V{r1}.\V{a}_\K{AND}_\V{r1}.\V{b}=\V{t3}.\V{b}}}
\end{lgrind}
\vspace{2mm}

This query is intercepted and rewritten by Verity as:

\vspace{2mm}
\begin{lgrind}
\L{\LB{\K{SELECT}_\V{t1}.\V{x},_\V{t1}.\V{y},_\V{t1}.\V{a},}}
\L{\LB{}\Tab{11}{\V{r1}.\V{a},_\V{r1}.\V{s},_\V{r1}.\V{b},}}
\L{\LB{}\Tab{11}{\V{t3}.\V{c},_\V{t3},\V{d},_\V{t3},\V{b}}}
\L{\LB{\K{FROM}_\V{t1},_(\K{SELECT}_\V{a},_\V{s},_\V{b}_\K{FROM}_\V{t2})_\K{AS}_\V{r1},_\V{t3}}}
\L{\LB{\K{WHERE}_\V{t1}.\V{a}=\V{r1}.\V{a}_\K{AND}_\V{r1}.\V{b}=\V{t3}.\V{b}}}
\end{lgrind}
\vspace{2mm}

\begin{algorithm2e}[t]
	\SetAlgoLined
	\tcp{$Q_{parent}$ is empty for the outermost query}
	\If{$Q_{parent}$ == $\phi$}{ 
		$Q_{orig}$ $\gets$ Q\; \label{alg:select:qorig}
	}
	\BlankLine
	\If{Q has nested SELECT queries}{ \label{alg:select:recurse-begin}
	  \tcp{Lookup only one level lower SELECT queries}
	  \ForEach{Q' in next level SELECT queries}{
		\tcp{Call \Select\ recursively on the subqueries
		with the parent query}
		\Select(Q', Q)\;
	  }
	} \label{alg:select:recurse-end}
	\BlankLine
	\tcp{In place projected attr change}
	Q $\gets$ ChangeProjection(Q)\; \label{alg:select:changeproj}
  \BlankLine
	\If{$Q_{parent}$ == $\phi$}{
	  \tcp{table\_list can be populated recursively while parsing, and these can
	  include temporary tables as shown in our example}
	  \textit{table\_list} $\gets$  tables in Q\;
	  $Res_{Q}$ $\gets$ \dbexecute(Q)\; \label{alg:select:dbexec}
	  \BlankLine
	  \ForEach{T in table\_list}{ \label{alg:select:foreachtable-begin}
		\tcp{Project out only attributes of table T from the whole
		result row}
	 	tuples $\gets$ TuplesOf($Res_{Q}$, T)\; 
		\For{r $\in$ tuples}{
			$\mathbf{RowID}_{r,T} = {\sf h}(PrimaryKey_{r}.T)$\;
			$\mathbf{fingerprint}_{r} = {\sf h}(\mathit{RowID}.r)$\;
			Verify \textit{fingerprint} against Hyperledger\;
		}
	  } \label{alg:select:foreachtable-end}
	  \BlankLine
	  \tcp{Get results with original projection attributes}
	  $Res_{Q_{orig}}$ $\gets$ ProjectResults($Res_{Q}$, $Q_{orig}$)\; \label{alg:select:origres}
	  \Return $Res_{Q_{orig}}$\;
	}
	 \caption{\Select(Q, $Q_{parent}$)} \label{alg:select}
\end{algorithm2e}

Algorithm \ref{alg:select} shows programmatic way of doing this procedure for a SELECT query having any level of nesting.
At the beginning Algorithm \ref{alg:select} receives the original user query $Q$, with $Q_{parent}$ empty. The original query is stored
for later reference as $Q_{orig}$ (line \ref{alg:select:qorig}). If the query has nested SELECT queries, then they are parsed
recursively. Every time \Select~is invoked on a nested query, the outer SELECT query is sent as $Q_{parent}$. Note here that since 
this is a bottom-up parser, the outermost query does not get parsed or processed until all the inner queries are done parsing (lines \ref{alg:select:recurse-begin}--\ref{alg:select:recurse-end}).

When the query has no more nested subqueries, we change the projected attributes in the current query to include all the attributes
of all the base tables in that query (line \ref{alg:select:changeproj}). Recall that we store the DBMS schema with Verity 
to be able to do this. After the innermost subquery is parsed, recursion
unwinds, and all the parent queries are processed to change projected attributes from their respective base tables.
When the recursion ends, we are back to the outermost query. Here we get a new $Q$ different from $Q_{orig}$ which projects
out all the attributes in the accessed base tables. Then this query $Q$ is executed on the DBMS (line 
\ref{alg:select:dbexec}). \textit{table\_list} contains all the base tables accessed in the query, including any subqueries.
\textit{table\_list} is populated while the queries are parsed recursively in a bottom-up manner. This step is not shown
explicitly in the algorithm for simplicity of presentation.

The results returned by this modified $Q$ are cached in $Res_Q$. Note that this is a temporary caching of the results, and Verity does 
not store $Res_Q$ persistently in any manner. Then for each unique base table in the query, we project out
tuples of only that table from $Res_Q$ by using the DBMS schema stored in Verity, and verify the tuple fingerprint against
Hyperledger Fabric (lines \ref{alg:select:foreachtable-begin}--\ref{alg:select:foreachtable-end}). Once this verification is done,
we project out the attributes from $Res_Q$ according to the original query $Q_{orig}$, and generate the results of the original user query (line \ref{alg:select:origres}).

\subsubsection{UPDATE queries}
\label{sec:update}
The grammar for UPDATE queries is as given below.
\begin{verbatim}
update_statement: `UPDATE' single_table `SET' 
                set_clauses [`WHERE' qualifications]
set_clauses: set_clause `,' set_clauses
            | set_clause
set_clause: identifier `=' expression
           | identifier `=' `(' select_statement `)'
\end{verbatim}

An UPDATE query can have nested SELECT queries as a part of the SET clause. For instance consider the following UPDATE query.

\vspace{2mm}
\begin{lgrind}
\L{\LB{\V{UPDATE}_\V{T1}_\V{SET}_\V{T1}.\V{a}_=_}}
\L{\LB{(\K{SELECT}_\V{a}_\K{FROM}_\V{T2}_\K{WHERE}_\V{T2}.\V{key}=\N{1234})}}
\L{\LB{\K{WHERE}_\V{T1}.\V{b}_=_\N{4567}}}
\end{lgrind}
\vspace{2mm}

In this query whichever row of $T1$ has ``b'' column with value \textit{4567}, its respective ``a'' column is updated with
a value of ``a'' from $T2$ such that $T2.key$ in the same row has value \textit{1234}. Note that in this query the nested SELECT 
query must always return a \textit{unique} single value, and not a list of values (as that will violate the arithmetic of ``='' operator).

Here Verity uses the same principle of modifying the query along with its nested queries
to get -- (a) the entire tuples that are going to provide update values, (b)  old tuples that are going to get updated. It first
checks the fingerprints of all these rows, generates new fingerprints for the updated rows, stores them on Hyperledger,
and then sends the updated rows to the DBMS. This is achieved methodically using Algorithm \ref{alg:update}. Its functioning
is explained briefly as follows.

\begin{algorithm2e}
\SetAlgoLined
rows[] $\gets \phi$\;

\If{SET clause has SELECT queries}{ \label{alg:update:each-select-start}
	\ForEach{SELECT query Q'}{
	\tcp{Invoke Algorithm \ref{alg:select}}
	  rows[Q'] $\gets$ \Select(Q', $\phi$)\tcp*{verify fingerprints}
	}
}\label{alg:update:each-select-end}
\BlankLine
$T \gets$ get table to be updated from Q\; \label{alg:update:newq-begin}
\textit{qualifications} $\gets$ get \texttt{WHERE} clause of Q\;
Q'' $\gets$ ``\texttt{SELECT * FROM $T$ WHERE qualifications}''\; \label{alg:update:newq-end}
old\_rows $\gets$ \dbexecute(Q'')\; \label{alg:update:dbexec}
Check fingerprint of the old rows\; \label{alg:update:checkoldkeys}
\BlankLine
updated\_rows $\gets$ Construct(Q, rows, old\_rows)\; \label{alg:update:construct}
\ForEach{r $\in$  updated\_rows}{ \label{alg:update:foreachrow-start} 
            $\mathbf{RowID}_{r,T} = {\sf h}(PrimaryKey_{r}.T)$\;
         $\mathbf{fingerprint}_r = {\sf h}(\mathit{RowID}.r)$\;
         Send update to Hyperledger\;
         Send individual row update to SQLite\; \label{alg:update:sendupdatetodb}
} \label{alg:update:foreachrow-end}
 \caption{\Update(Q)} \label{alg:update}
\end{algorithm2e}

If the SET clause of UPDATE query has nested SELECT queries, then for each such SELECT query
$Q'$, we invoke Algorithm \ref{alg:select} for \Select~ to get the tuples used in UPDATE. Note that 
Algorithm \ref{alg:select} also methodically checks the integrity of the tuples accessed by
this SELECT query, thus ensuring that the tuples used for an update have not been tampered.
The SET clause can have multiple such SELECT queries for each set condition. The results returned
for each SELECT query $Q'$ are stored separately as \textit{rows[Q']} (lines \ref{alg:update:each-select-start}--\ref{alg:update:each-select-end}). 
Then if there is a WHERE clause in the original UPDATE query, e.g., \texttt{WHERE T1.b = 4567} of the example query given above, we
retrieve old tuples from table $T$ that will get updated (lines \ref{alg:update:newq-begin}--\ref{alg:update:dbexec}).

We check the fingerprints of these tuples against Hyperledger (line \ref{alg:update:checkoldkeys}).
Next we \textit{construct} the new tuples that will be in the table  after the update (line \ref{alg:update:construct}),
create their fingerprints, and send them to Hyperledger.
Note that since we have completely deconstructed the original UPDATE query and have 
individual updated tuples, we simply send these individual tuples to the DBMS than the original UPDATE query $Q$ (lines \ref{alg:update:foreachrow-start}--\ref{alg:update:foreachrow-end}).

\subsubsection{INSERT queries} \label{sec:insert}

SQL grammar for an INSERT query is as given below.
\begin{verbatim}
insert_statement: `INSERT INTO' single_table 
        `(' id_list `)' `VALUES' `(' expr_list `)' 
        | `INSERT INTO' single_table 
        `(' id_list `)' select_statement 
\end{verbatim}

\begin{algorithm2e}
\SetAlgoLined
$T \gets$ get table to be inserted in from $Q$\;
\If{Q has nested SELECTs}{ \label{alg:insert:select-start}
 $Q'$ $\gets$ nested SELECT query\;
 \tcp{Invoke Algorithm \ref{alg:select} and verify tuples}
  rows $\gets$ \Select($Q'$, $\phi$)
}\label{alg:insert:select-end}
\Else{ 
  rows $\gets$ Populate rows from Q\; \label{alg:insert:nonesting}
}
\For{r $\in$  rows}{ \label{alg:insert:foreachrow-start}
    $\mathbf{RowID}_{r,T} = {\sf h}(PrimaryKey_{r}.T)$\;
    $\mathbf{fingerprint}_r = {\sf h}(\mathit{RowID}.r)$\;
    Insert fingerprint in Hyperledger\;
	Insert tuple in SQLite;
}\label{alg:insert:foreachrow-end}
 \caption{\Insert(Q)} \label{alg:insert}
\end{algorithm2e}

Like UPDATE, INSERT query too can have nested SELECT statements in it. This is  when the tuples
are inserted by constructing them out of a results of another SELECT query.
We have given a methodical way of handling an INSERT query in Algorithm \ref{alg:insert}.
If the INSERT query has a nested SELECT, we invoke Algorithm \ref{alg:select}, verify the 
tuples accessed by this SELECT query, done as a part of Algorithm \ref{alg:select}, and get the tuples 
to be inserted (lines \ref{alg:insert:select-start}--\ref{alg:insert:select-end} in Algorithm 
\ref{alg:insert}). Else tuples are populated from the original query $Q$ (line 
\ref{alg:insert:nonesting}). For each new tuple to be inserted, we generate its fingerprint
and store it on Hyperledger. Like UPDATE queries, here too we deconstruct the INSERT
query completely to generate each new tuple, and thus we send these individual tuples to the DBMS
for insertion than executing the original INSERT query (lines \ref{alg:insert:foreachrow-start}--\ref{alg:insert:foreachrow-end}).

\subsubsection{DELETE queries}
\label{sec:delete}
DELETE query's grammar is as given below.

\begin{verbatim}
delete_statement: `DELETE FROM' single_table
                [`WHERE' qualifications]
\end{verbatim}

Processing of a DELETE query is very similar to an INSERT query. DELETE query does not contain
any nested SELECT query. This is because currently we do not process \texttt{IN, ANY, EXISTS, 
GROUP-BY, HAVING} clauses, which may in turn contain nested SELECTs as a part of the WHERE 
qualifications. Thus processing of DELETE queries just involves fetching the tuples to be deleted, verifying their fingerprints before deletion, marking their fingerprints as deleted on Hyperledger, and then deleting them from DBMS.
Algorithm \ref{alg:delete} shows this procedure.

\begin{algorithm2e}
\SetAlgoLined
$T \gets$ get table to be deleted from $Q$\;
\textit{qualifications} $\gets$ get WHERE clause of $Q$\;
\BlankLine
$Q' \gets$ ``\texttt{SELECT * FROM $T$ WHERE qualifications}''\;
\tcp{Verify rows to be deleted first}
rows $\gets$ \Select(Q', $\phi$)\;
\For{r $\in$  rows}{
    $\mathbf{RowID}_{r,T} = {\sf h}(PrimaryKey_{r}.T)$\;
    $\mathbf{fingerprint}_r = {\sf h}(\mathit{RowID}.r)$\;
	Mark fingerprint deleted on Hyperledger\;
	Delete $r$ from SQLite;
}
 \caption{\Delete(Q)} \label{alg:delete}
\end{algorithm2e}

\section{Discussion and Future Work} \label{sec:discussion}
In this section we discuss additional factors for future enhancements in Verity.

\subsection{Thwarting SQL Injection Attack} \label{sec:sqlinjection}
In Section \ref{sec:sql}, we elaborated on our main contribution, i.e., our way of intercepting SQL queries using the SQL grammar, 
methodically deconstructing each nested SQL query, and verifying the integrity of the tuples returned by that query.
A positive side effect of this is -- Verity's SQL parser can act as an intermediary that 
can thwart SQL injection attacks. In the future, Verity's SQL parser can also be empowered with 
checks for malicious SQL statements.

\subsection{ACID Properties} \label{sec:atomicity}
ACID (Atomicity, Consistency, Isolation, Durability) properties make the core of any mature DBMS. Verity
is a dataless framework between a blockchain and a DBMS (it does not store any data or metadata with itself), thus
durability does not apply to Verity. However, queries that modify the tuples in DBMS are processed via Verity,
and thus need atomicity, consistency, and isolation properties of transaction management.
As given in Section \ref{sec:sql}, Verity maintains information about the schema of the database. Using that,
in the future, we plan to have an elaborate method to handle concurrent SQL queries that are modifying tuple
values. DBMS transaction management is a well-studied topic \cite{dbbook}, and borrowing the same concepts of transaction
management and serializability, we can handle concurrent SQL queries modifying the tuples using the DBMS
schema.

\subsection{Optimizations} \label{sec:optimizations}

Currently the Verity SQL parser is single-threaded, processes one query at a time, and
makes blockchain fingerprint lookup one tuple at a time for that query. Also for any transactional queries (queries
that modify tuples), to maintain atomicity, consistency, and isolation (ACI properties) Verity immediately pushes
them on Hyperledger. The typical block creation process is time consuming, and Hyperledger Composer adds further overheads \cite{composerpoorperf}.
This is the reason for our relatively higher total query execution times as shown in Table 
\ref{table:result}, Figure \ref{fig:result}, and our supplementary report \cite{veritytechrep}.
In the future we will make tuple fingerprint lookups on Hyperledger in parallel to improve overall query processing time,
and investigate into the throughput of Hyperledger Fabric and Composer framework.
These optimizations will go hand-in-hand with the previousely mentioned aspects in Section \ref{sec:atomicity}.

SQL language is a well-researched topic with a rich set of literature for manipulating and rewriting SQL queries
for better performance. In the current Verity framework, we have adopted a way of projecting out all the columns
of the tables in a nested SELECT query for integrity checking. This is
due to the way tuple fingerprints are stored on the blockchain.
However in the future, we can have more intelligent logic based on the structure of the query, for parsing
and tuple checking, which can reduce the performance overhead and blockchain lookup time.

\subsection{Indexes, Views on Base Tables} \label{sec:indexviews}
Indexes and views on tables are common in modern DBMS. The use of indexes or views in query execution is most times
opaque to the end user and is decided internally by the DBMS query optimization method.
Thus in an insider attack, it is possible that the attacker modifies tuples in the base tables of a DBMS,
but indexes or views created on them are not updated. In such a case, the database itself 
is in \textit{inconsistent state}. Since Verity considers DBMS to be an independent entity, we assume that these 
consistency checks across base tables and their indexes or views will be done by the DBMS.
However, we want to highlight that the goal of Verity is to preclude illegitimately tampered tuples from being 
used in subsequent SQL queries, and avoid the cascade effect in important decisions based on them,
e.g., academic grades or financial entities. Verity allows any tuples retrieved from DBMS that pass the
integrity check against blockchain fingerprints. In case of any illegitimate modifications found in tuples,
it flags them, and prevents sending the results of the query.

\subsection{Special Case of Illegitimate Delete} \label{sec:discussdelete}
In Section \ref{sec:delete}, we outlined our procedure for handling a DELETE query that comes through the Verity framework.
However, an insider attack containing illegitimate tuple deletions presents a different challenge for detecting it
purely through the subsequent SQL queries and their results. If an inside attacker deletes tuples from DBMS (and erases
any logs relating to that too), the respective tuples never get selected as a part of any subsequent SQL queries,
never show up in Verity framework for checking against blockchain fingerprints, and thus deletion does not get detected.

We propose the following solution for this. Recall from Section \ref{sec:privacy}, that along with the fingerprint of each tuple, we 
also store the total number of tuples in a table on the blockchain. One solution for detecting illegitimate delete is to 
periodically run \texttt{SELECT count(*) FROM T} query for each table \texttt{T} in the DBMS, and cross verify the number of tuples
returned with the tuple count stored on the blockchain. However, this check can be fooled with an illegitimate delete followed
by a dummy (illegitimate) insert in the same table. To handle this case, Verity framework can periodically
run \texttt{SELECT * FROM T} query for every table \texttt{T}, and check every tuple's fingerprint against
that stored on the blockchain. The dummy tuple's fingerprint will not be found on the blockchain, and an alert will be raised for
tampering of tuples.

\section{Experiments}\label{sec:expt}

In this section we present our experiments using Verity.

\begin{figure*}[ht]
	
	\begin{subfigure}{0.5\textwidth}
		\includegraphics[width=\linewidth]{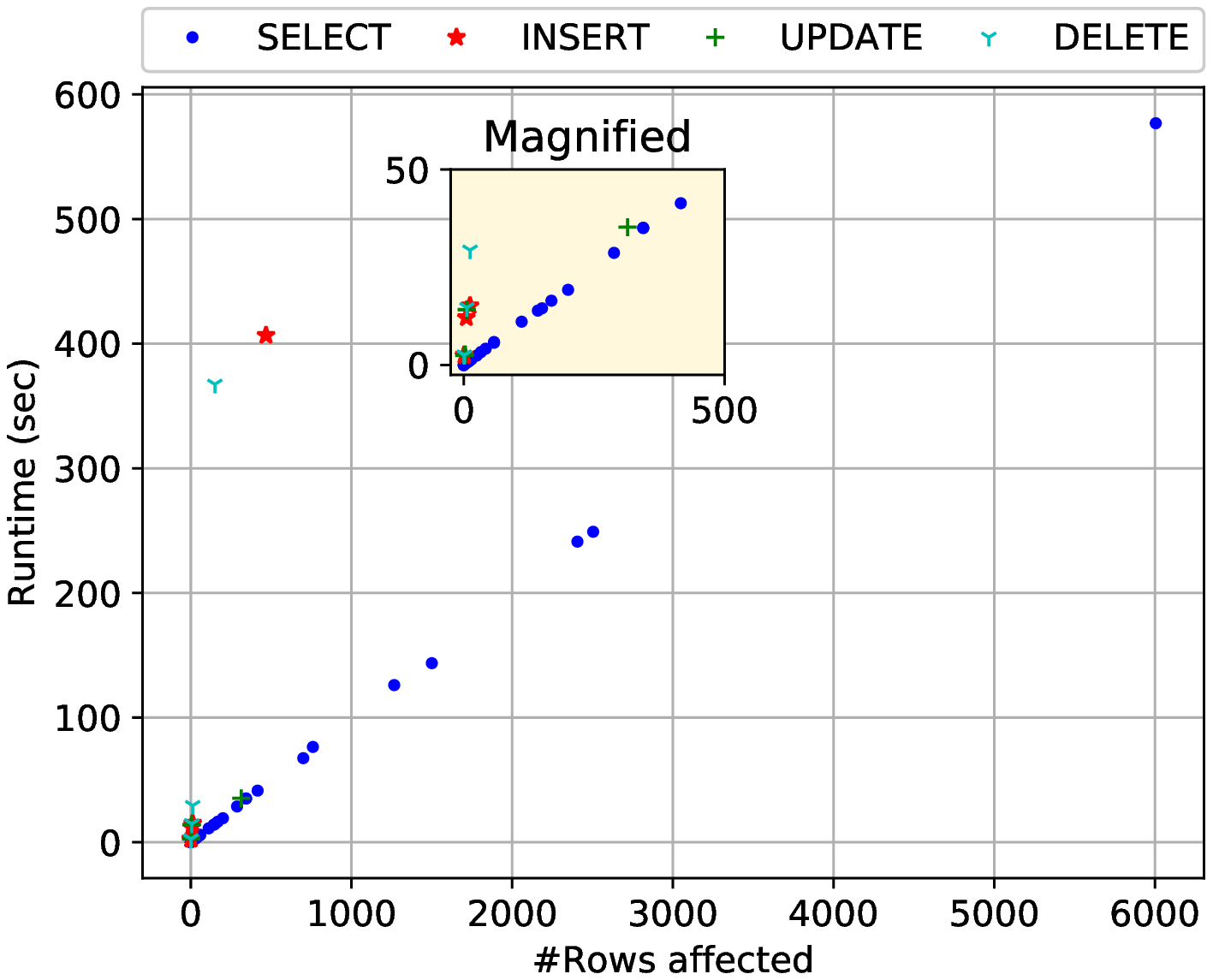} 
		\caption{{\it Scaling Factor=$0.001$}}
		\label{fig:subim1}
	\end{subfigure}
	\begin{subfigure}{0.5\textwidth}
		\includegraphics[width=\linewidth]{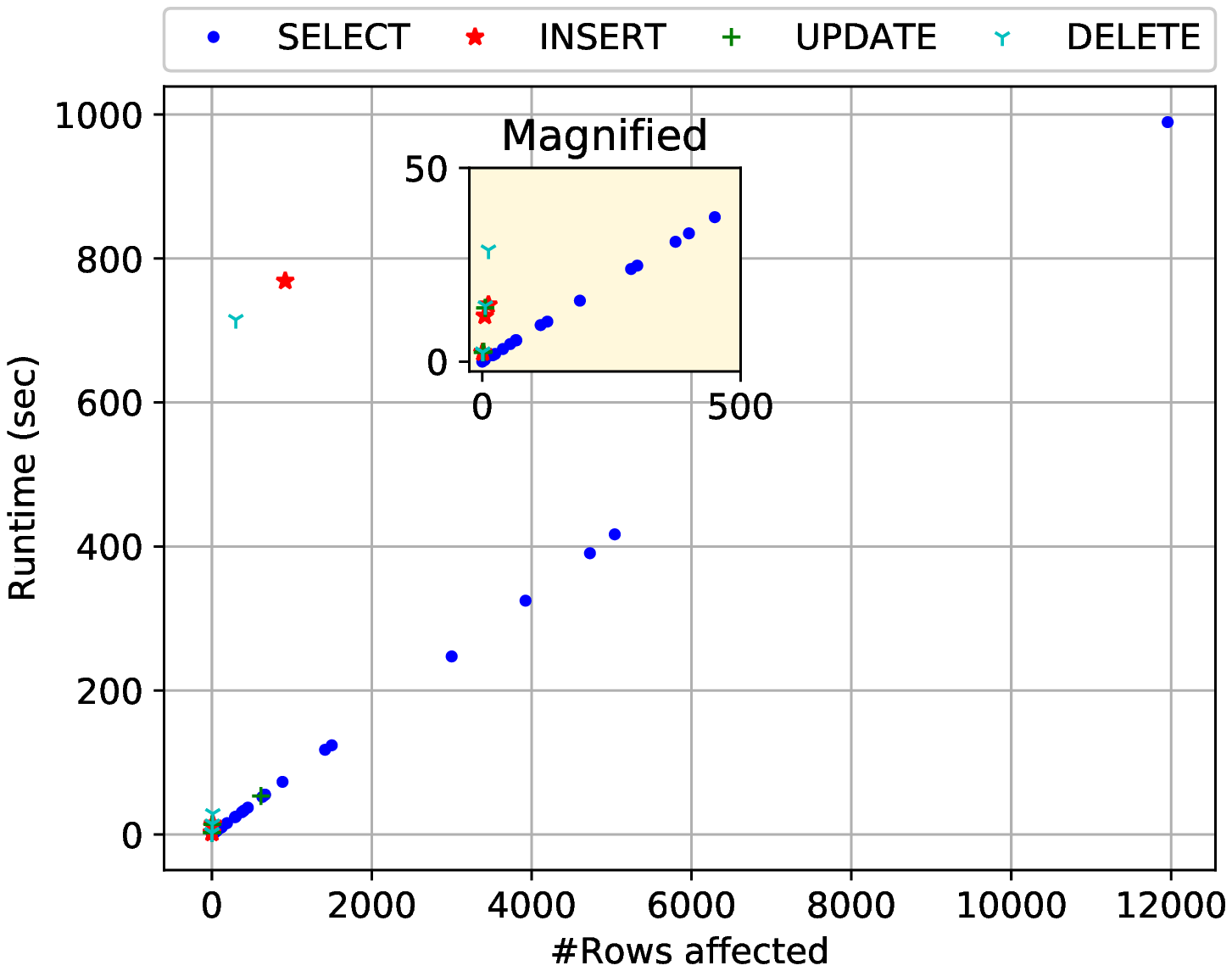}
		\caption{{\it Scaling Factor=$0.002$}}
		\label{fig:subim2}
	\end{subfigure}\\
	\begin{subfigure}{0.5\textwidth}
		\includegraphics[width=\linewidth]{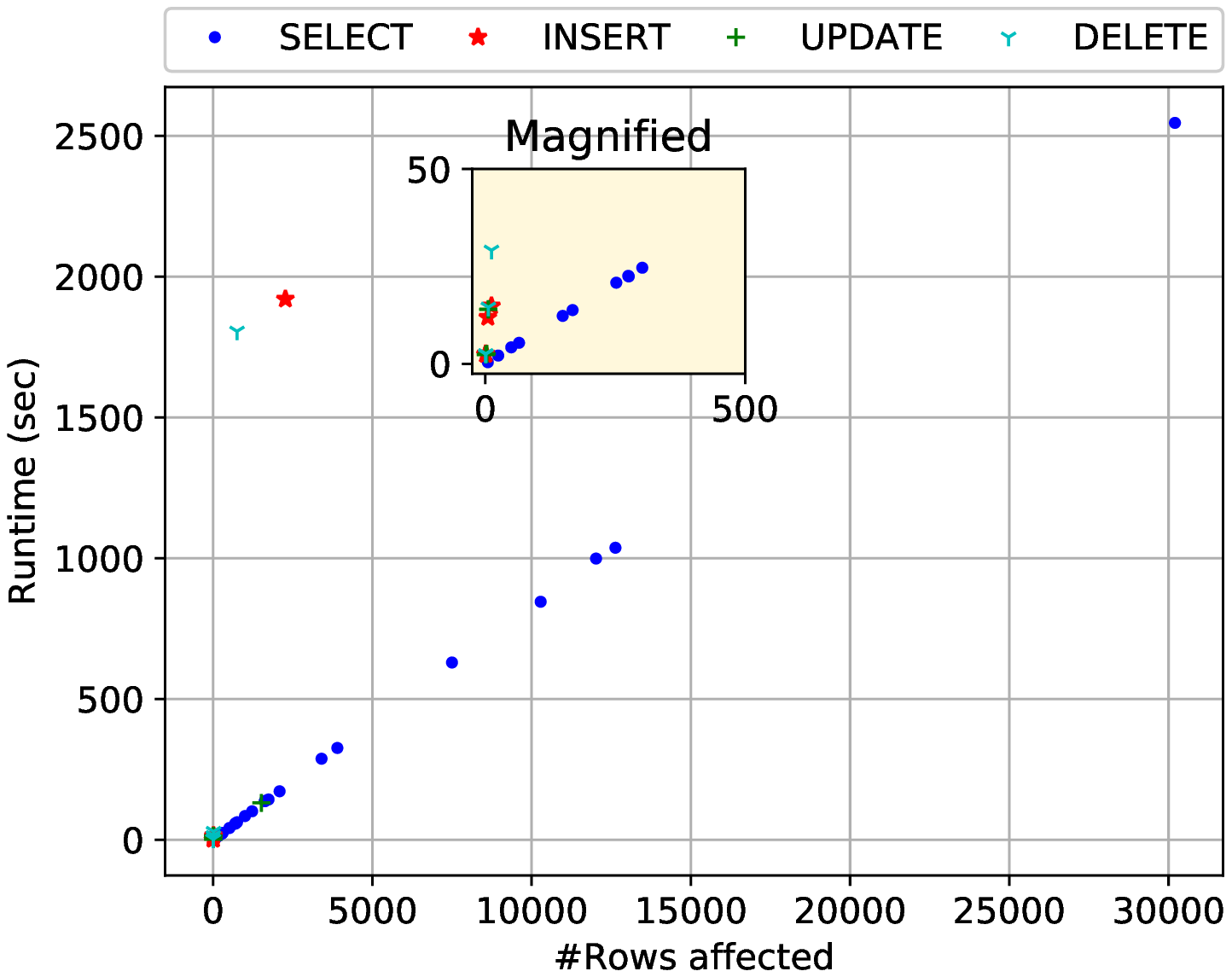} 
		\caption{{\it Scaling Factor=$0.005$}}
		\label{fig:subim1}
	\end{subfigure}
	\begin{subfigure}{0.5\textwidth}
		\includegraphics[width=\linewidth]{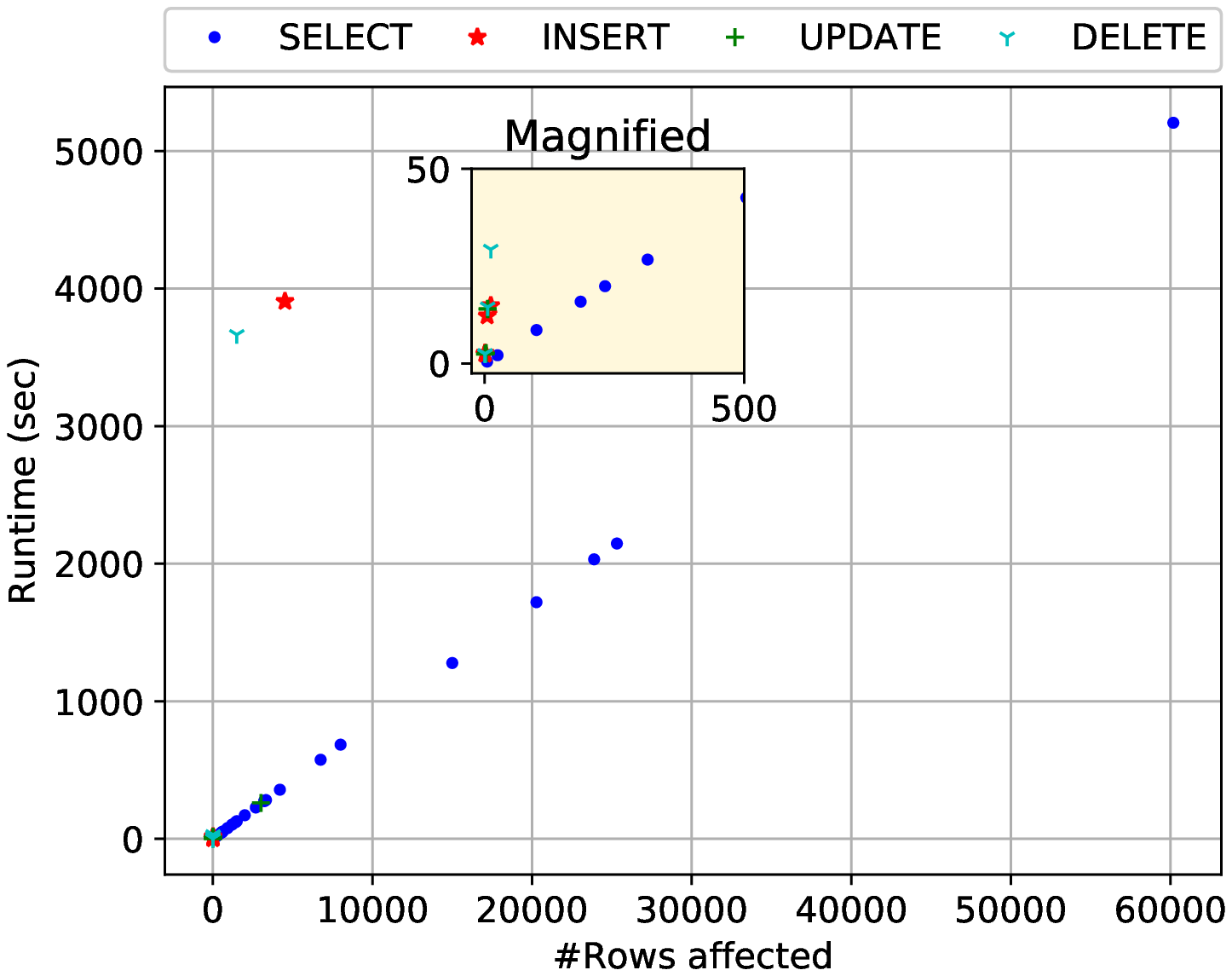}
		\caption{{\it Scaling Factor=$0.01$}}
		\label{fig:subim2}
	\end{subfigure}
	
	\caption{Plots of relation between processing time vs number of effective tuples processed as per our parsing method (see Section \ref{sec:sql}) for 4 TPC-H scaling factors, on 43 queries as given in Appendix \ref{appendix:sql}.}
	\label{fig:result}
\end{figure*}

\subsection{Computation Setup} \label{sec:setup}
As described before in Section \ref{sec:arch}, we use Hyperledger Fabric v1.1 as the blockchain network,
and Hyperledger Composer v0.19.18 REST API to write the chaincodes (for interacting with Hyperledger Fabric).
Hyperledger Fabric was deployed using docker version 18.09.0, build 4d60db4. 
We wrote our SQL parser using Python v3.5.2 programming language, and used SQLite v3.11.0 as the underlying DBMS.

For testing, we deployed this setup on two Asus 2U servers with Intel C602-A chipset, Intel Core Xeon E5-2620 processor, 
\done\todo{should we mention number of cores on openstack vm?} 16GB RAM  and 1TB SATA Hard disk running Ubuntu 16.04.01 with Linux 4.4.0-62-generic kernel. We used one machine to deploy Hyperledger Fabric, and used another to parse the queries and send requests to Hyperledger. The machines were networked together using a CAT5 LAN cable. \done\todo{Which machine SQLite deployed on?}.

\subsection{Dataset and SQL Queries}
We used the TPC-H \cite{tpch} synthetic relational data generator in our experiments, and generated four datasets using scaling factors 
0.001, 0.002, 0.005, and 0.01.
Note that, although we have used TPC-H, our goal in Verity framework is \textit{not} to benchmark SQLite or Hyperledger Fabric,
but to demonstrate functioning of Verity under variety of SQL queries.
Table \ref{table:num_rows} shows the number of tuples in each table within these datasets corresponding to each scaling factor.

Using the 22 queries from TPC-H suite as guidelines, we generated 43 queries of varying
complexity and intermix of SELECT, INSERT, UPDATE, and DELETE\footnote{\small{TPC-H suite did not have many update queries.}}. All 
the 43 queries are given in Appendix \ref{appendix:sql}. In summary -- we used 27 SELECT queries having nested subqueries as well as 
joins, 6 INSERT, 5 UPDATE, and 5 DELETE queries, simple as well as with nested SELECT statements within them. Note that the queries have 
varying levels of complexity and nested structures within them.

\begin{table}[h]
\centering 
\begin{tabular}{ |p{2cm}||p{1cm}|p{1cm}|p{1cm}|p{1cm}|  }
 \hline
 & \multicolumn{4}{c|}{\textit{Number of tuples}} \\
\hline
\textit{Scale Factors} & \textit{0.001} & \textit{0.002} & \textit{0.005} & \textit{0.01}\\
 \hline
 Customer & 150 & 300 & 750 & 1500 \\
 Lineitem & 6005 & 11957 & 30201 & 60175  \\
 Nation & 25 & 25 & 25 & 25 \\
 Orders & 1500 & 3000 & 7500& 15000  \\
 Part & 200 & 400 & 1000 & 2000 \\
 Patsupp & 700 &1500 & 3900 & 8000  \\
 Region & 5 & 5 & 5 & 5 \\
 Supplier & 10 & 20 & 50 & 100 \\
 \hline
 \hline
 \textit{Total} & 8595 & 17207 & 43431 & 86805\\
 \hline
\end{tabular}
\caption{Number of tuples in each table of the TPC-H dataset}\label{table:num_rows}
\end{table}

\begin{table}[t]
	\begin{tabular}{|L{0.4cm}|L{0.9cm}|C{1cm}|C{0.7cm}|L{0.9cm}|L{1.2cm}|L{0.7cm}|}

		\hline
		\textit{Q No.}  & \textit{Q Type} & \textit{Is nested query?} & \textit{\# Tables} & \textit{\# Tuples} & \textit{End-to-end time} & \textit{Time per tuple} \\*  \hline
		
		\hline\hline\multicolumn{7}{|c|}{\it Scaling Factor = 0.001}\\ \hline
		2  & S  & \xmark & 1 & 5    & 0.52   & 0.10 \\ \hline
		11 & U(S) & \cmark & 3 & 3    & 2.68   & 0.89 \\ \hline
		18 & I  & \xmark & 1 & 5    & 12.12  & 2.42 \\ \hline
		19 & U  & \xmark & 1 & 6    & 14.18  & 2.36 \\ \hline		
		26 & D  & \xmark & 1 & 12   & 29.37  & 2.45 \\ \hline
		40 & S(S)  & \cmark & 3 & 2504 & 249.08 & 0.10 \\ \hline
		
		\hline\hline\multicolumn{7}{|c|}{\it Scaling Factor = 0.002}\\ \hline
		2  & S  & \xmark & 1 & 5    & 0.48   & 0.10 \\ \hline
		11 & U(S) & \cmark & 3 & 3    & 2.55   & 0.85 \\ \hline
		18 & I  & \xmark & 1 & 5    & 11.84  & 2.37 \\ \hline
		19 & U  & \xmark & 1 & 6    & 13.87  & 2.31 \\ \hline		
		26 & D  & \xmark & 1 & 12   & 28.80  & 2.40 \\ \hline
		40 & S(S)  & \cmark & 3 & 5040 & 416.84 & 0.08 \\ \hline
		
		\hline\hline\multicolumn{7}{|c|}{\it Scaling Factor = 0.005}\\ \hline
		2  & S  & \xmark & 1 & 5     & 0.45     & 0.09 \\ \hline
		11 & U(S) & \cmark & 3 & 3     & 2.59     & 0.86 \\ \hline
		18 & I  & \xmark & 1 & 5     & 11.97    & 2.39 \\ \hline
		19 & U  & \xmark & 1 & 6     & 13.96    & 2.33 \\ \hline
		26 & D  & \xmark & 1 & 12    & 29.14    & 2.43 \\ \hline
		40 & S(S)  & \cmark & 3 & 12632 & 1,037.30 & 0.08 \\ \hline		
		
		\hline\hline\multicolumn{7}{|c|}{\it Scaling Factor = 0.01}\\ \hline
		2  & S  & \xmark & 1 & 5     & 0.52     & 0.10 \\ \hline
		11 & U(S) & \cmark & 3 & 3     & 2.66     & 0.89 \\ \hline
		18 & I  & \xmark & 1 & 5     & 12.25    & 2.45 \\ \hline
		19 & U  & \xmark & 1 & 6     & 14.08    & 2.35 \\ \hline
		26 & D  & \xmark & 1 & 12    & 29.30    & 2.44 \\ \hline
		40 & S(S)  & \cmark & 3 & 25314 & 2,146.84 & 0.08 \\ \hline
		
\end{tabular}

\caption[Selected Query Stats]{Execution stats of $6$ representative queries from full report \cite{veritytechrep}.
``QNo.'' as per list in Appendix \ref{appendix:sql}. ``\# Tables'' -- tables accessed by the query, ``\# Tuples'' -- total number of tuples accessed/verified. S, I, U, D stands for SELECT, INSERT, UPDATE, DELETE respectively, letter in bracket shows type of nested query.} \label{table:result}
\end{table}

\subsection{Performance Metrics}
For experiments we used the following metrics:

\begin{enumerate}[noitemsep]
\item TPC-H scaling factor, 0.001, 0.002, 0.005, 0.01.

\item Time taken for \textit{end-to-end} execution of each query, i.e., wall-time, averaged over 5 runs.

\item Time taken for the execution of SQL queries on SQLite (Section \ref{sec:sql} elaborates
how the queries are executed).

\item Time taken for Hyperledger Fabric lookup of fingerprints of the tuples in the query results (Section \ref{sec:sql}
describes how tuples are extracted for lookup from the SQL query results).

\item Number of tuples accessed, inserted, or modified by a query -- in case of an INSERT, UPDATE, DELETE query with nested
SELECT subquery, this number includes the tuples  accessed by the nested SELECT query as well as the ones modified by the INSERT,
UPDATE, or DELETE query.

\end{enumerate}

\subsection{Analysis of the Results} \label{sec:exptanalysis}
We ran 43 queries on the data generated by each TPC-H scaling factor separately, and noted the \textit{end-to-end} wall time,
time taken to run the queries on SQLite, and time taken by Hyperledger Fabric averaged over 5 runs. 
We observed that the time taken for running the queries on SQLite was very negligible compared to the Hyperledger lookup.
Hence we plotted the end-to-end runtime of the 43 queries against the total number of tuples affected or accessed in each 
query for each scaling factor. Figure \ref{fig:result} shows these plots.

From the plots in Figure \ref{fig:result}, we can clearly see a linear relationship between the number of tuples affected
by a query and total query runtime. Recall from our architecture given in Section \ref{sec:arch}, that Verity acts only as 
a framework facilitating the use of a blockchain network with an SQL DBMS. As given in
Section \ref{sec:metadata}, we store only \textit{256 bit fixed-length} fingerprint of each tuple on the blockchain 
network. Thus lookup overhead per tuple remains constant, and scales linearly as per total number of tuples.

For an interested reader, we have also given the exact execution times for each query, for each scaling factor in our supplementary report at \cite{veritytechrep}. In Table 
\ref{table:result}, we present the execution statistics for $6$ representative queries from this report of varying types (S, 
I, U, D), complexity, and nesting structure for quick insights into the results. From this detailed analysis, it can be 
observed that for a SELECT query, typical Hyperledger lookup time is 0.08--0.1 second per tuple, and for queries requiring 
modification of tuples, such as INSERT, UPDATE, DELETE, this time is about 0.8--2.5 seconds per tuple. The higher time for 
the queries requiring modifications in the tuples is due to the peer-consensus protocol that needs to add a
fingerprint, or adjust the number of tuples per table on the Hyperledger network (ref. Sections \ref{sec:metadata}, \ref{sec:privacy}).

From these results, one may feel that use of blockchains along with a DBMS incurs overheads that seem 
undesirable for high-performance throughput. However, as discussed before in the text of this paper, in the present
work, our aim is to -- (a) establish a formalism for handling a rich set of complex SQL queries,
(b) without migrating entire DBMS data on a blockchain, (c) thereby maintaining privacy, (d) while using the tamper-resistance properties of a blockchain to detect an insider attack.
We have achieved this by intercepting SQL queries and their resulting tuples, by treating both
the blockchain as well as DBMS as black-boxes. Verity itself does not do any SQL query optimization, or
data or metadata caching and indexing. Thus the overall throughput of the Verity framework can be improved by 
investigating the blockchain throughput improvement methods (which is not the focus of our present work). Nevertheless, in Section \ref{sec:optimizations} we have discussed the performance optimizations we intend to do in the Verity framework
through parallelizing the blockchain lookups, and investigating SQL query structures.

\section{Related Work} \label{RelatedWork}

We classify related work broadly into categories given in the following subsections.

\subsection{Blockchains and DBMS}
As discussed briefly in Section \ref{intro}, current efforts for using blockchain technology 
for DBMS mainly offer solutions for integrating blockchain functionalities, such as peer-consensus protocol
and decentralized storage into native DBMS. Big-chainDB \cite{bigchaindb18} integrates Tendermint 
\cite{tendermint18,tendermint14} with MongoDB \cite{mongodb18}. It supports only decentralized blockchain based 
data management eco-system, and only MongoDB's querying interface. LedgerDB \cite{ledgerdb} supports high transaction 
throughput, but provides only a single table, and does not support various SQL features. ChainDB by Bitpay Inc 
\cite{chaindb} focuses on bitcoin transactions, than providing a general purpose DBMS solution.
EthernityDB \cite{EthernityDB} integrates a DBMS functionality into an Ethereum blockchain, by keeping all the data
on the chain, and by mapping DBMS functionalities onto Ethereum smart contracts.

In most of these solutions the data from native DBMS has to be migrated entirely onto a new 
blockchain powered system. Solutions such as BigchainDB, LedgerDB, or ChainDB do not provide rich SQL interface.
In comparison to them, our goal in Verity is to use blockchain functionality of non-repudiation with 
a native DBMS without complete migration of the data, and by just intercepting the SQL queries and their results for insider 
attack detection. In Verity, we store only \textit{fingerprints} of the tuples on a blockchain, thus maintaining
data privacy.

\subsection{Intrusion Detection Systems}

Intrusion Detection Systems (IDS) mainly use machine learning techniques to model legitimate or illegitimate
behaviour to detect an anomaly. Their performance heavily relies on the training data used for
modelling user behaviour \cite{4547660,4267567,Kandias10,Panigrahi2013,4024507}.
An interested reader can refer to a more comprehensive survey given in  \cite{insider18}.

IDSs specific for DBMS model access patterns of DBMS users and learn data dependencies among
data items. DEMIDS (DEtection of MIsuse in Database System) \cite{demids} analyzes access patterns of users by using audit 
logs, builds user profiles, and use these to detect unusual behaviour. RBAC (Role Based Access Control) 
\cite{bertino05} improves upon DEMIDS by creating profiles for each role instead of individual user and associating users 
with roles, which enables them to handle a large number of database users. They use Na{\"i}ve Bayes Classifier.
Wu et al \cite{Wu2009} do role profiling with role hierarchies. 
Another technique mines the dependencies among data items \cite{hu04}, to detect what 
set of data items are usually accessed before a particular data is changed. Any transactions not compliant with these 
pre-built models are treated as malicious.
WDDRM (Weighted Data Dependency Rule Miner) \cite{srivastava06} improves upon this technique by associating weight to an 
attribute based on the sensitivity of the data stored in that attribute. 
There are other techniques that use time signatures \cite{lee00, hashemi08} and Hidden Markov Models 
\cite{barbara02}.

\subsection{Mitigation and Prevention Techniques}

In \cite{Vance:2015:IAT:2871313.2871318}, Vance et al propose to make the user accountable for policy 
violation, and in \cite{Bishop2008}, Bishop et al propose attribute-based group access control (ABGAC) policies to act as a deterrent.
Wu et al \cite{Wu2011} propose the concept of ``active data leak prevention'' where they use an encrypted secure data 
container (SDC) to ensure that only authorized users are able to access the data in a trusted environment.  In \cite{Pramanik2004}, Pramanik et al propose policies that prohibit modification of a sensitive file, while 
another ``inappropriate" file is open. Chagarlamudi at al \cite{Chagarlamudi2009} propose sequential access checking 
technique that prevents execution of malicious user tasks by using PetriNets.
\textit{Confidentiality via camouflage} performs deterministic numerical interval-based responses of ad-hoc
queries to a DBMS \cite{GopalGG02,Garfinkel2002,Gopal2006}. 

Along with these techniques, access control policies are used to prevent insider attacks.
In \cite{GusJabbour2010StoppingTI}, they integrate security policy mechanism in the system. 
Other techniques augment access control with role based policies, 
trustworthiness of the users, and risk assessment of the roles \cite{Crampton2010TowardsAA,Baracaldo2012}. 
In \textit{Cyber Deception} \cite{cyberDeception}, they propose a technique based on software defined networks (SDN) to defend the network with extensive scrutiny of the network.

In summary, Verity's approach to insider attack detection is significantly different from IDS, or previously proposed prevention and 
mitigation techniques.
In Verity, we use a technique to detect any tuple with illegitimate modifications, that is accessed in subsequent SQL queries, through fingerprint checks against Hyperledger Fabric.
At the time of bootstrapping, Verity stores fingerprints of the tuples on Hyperledger, but that does not involve 
any machine learning of the system\footnote{\small{Special case of illegitimate deletion of tuples is discussed in Section \ref{sec:discussdelete}.}}. Also through Verity, one can use any blockchain network with any SQL DBMS by using 
their respective interfaces, and peer consensus protocols.

\section{Conclusion}\label{conclusion}

In this paper, we propose Verity -- a framework to use blockchains to detect an insider 
attack or data tampering in a DBMS. We have achieved this by storing 
\textit{fingerprints} of DBMS tuples on a blockchain (Section \ref{sec:metadata}), intercepting SQL queries (Section \ref{sec:sql}),
and verifying validity of the tuples accessed or affected by these queries against the blockchain fingerprints.
For any queries modifying the data that are processed through Verity, it can attribute the changes to the corresponding owner.
Verity has some latency due to its tuple integrity checks against blockchain for each query.
However, we trust that in systems where integrity of the DBMS tuples is 
important from the point of critical decisions based on them, this latency can be tolerated instead of compromising
the integrity of the data. Nevertheless, we have discussed ways of improving throughput in Section \ref{sec:optimizations}.
Verity framework facilitates use of blockchain's non-repudiation
along with a native SQL DBMS without the need of data migration or adoption of different query interface.
Verity framework does not store any data or metadata within itself persistently.
This makes Verity quickly adoptable in critical applications requiring data integrity, provided
some latency can be tolerated.
We have implemented Verity for an academic grade management system and
our current solution works very well there as the database does not have very high transaction rate and can tolerate
delays caused by the intervention of the blockchain.

\begin{appendices} 
\section{SQL {Queries}}\label{appendix:sql}

\noindent{\bf Q1}: \selectq{select} o\_year, nation, ( sum (volume) as mkt\_share) from (( select (o\_orderdate as 
o\_year), ((l\_extendedprice * (1 - l\_discount)) as volume), (n2.n\_name as nation) from part, supplier, lineitem, 
orders, customer, (nation as n1), (nation as n2), region where p\_partkey = l\_partkey and s\_suppkey = l\_suppkey and 
l\_orderkey = o\_orderkey and o\_custkey = c\_custkey and c\_nationkey = n1.n\_nationkey and n1.n\_regionkey = 
r\_regionkey and r\_name = ``asia'' and s\_nationkey = n2.n\_nationkey and o\_orderdate $>$ '1995-01-01' and 
o\_orderdate $<$ `1996-12-31' and p\_type = `large plated tin' ) as all\_nations);

\noindent{\bf Q2}: \selectq{select} * from region;

\noindent{\bf Q3}: \selectq{select} * from supplier;

\noindent{\bf Q4}: \selectq{select} * from nation;

\noindent{\bf Q5}: \updateq{update} customer set c\_name = ``sjadfd'', c\_address = ``kafawehrnj'', 
c\_phone=``7894561265'', 
c\_acctbal = 22, c\_mktsegment = ``klasjfaw'', c\_comment=``laksfnwe'' where c\_custkey = 91639739;

\noindent{\bf Q6}: \insertq{insert} into nation (n\_nationkey, n\_name, n\_regionkey, n\_comment) values ( 93793619 
,``algeria'', 123454556741 ,``haggle detect slyly agai'');

\noindent{\bf Q7}: \insertq{insert} into customer ( c\_custkey  , c\_name , c\_address , c\_nationkey , c\_phone , 
c\_acctbal ,c\_mktsegment, c\_comment )  values (91639739 , ``loren'', ``lipsum'', 93793619 , ``1234'', 234, 
``muspil'', ``nerol'');

\noindent{\bf Q8}: \insertq{insert} into region (r\_regionkey, r\_name, r\_comment) values (123454556741, 
``sambhal'', 
``jhfasfhf kajhfawerb idauhfwerbe aksfhnwejrb'');

\noindent{\bf Q9}: \deleteq{delete} from nation where n\_nationkey = 93793619;

\noindent{\bf Q10}: \deleteq{delete} from region where r\_regionkey = 123454556741;

\noindent{\bf Q11}: \updateq{update} supplier set s\_nationkey = (select c\_nationkey from customer where c\_custkey= 
91639739), s\_phone = ( select c\_phone from customer where c\_custkey= 91639739), s\_comment =``askdenrjuhereu'', 
s\_acctbal = 2  + 10   where s\_suppkey = 91639739 +1000;

\noindent{\bf Q12}: \selectq{select} s\_acctbal, s\_name, n\_name, p\_partkey, p\_mfgr, s\_address, s\_phone, 
s\_comment from part, supplier, partsupp, nation, region where p\_partkey = ps\_partkey and s\_suppkey = ps\_suppkey 
and 
p\_size = 1 and s\_nationkey = n\_nationkey and n\_regionkey = r\_regionkey and r\_name = ``africa'';

\noindent{\bf Q13}: \selectq{select} n\_name, (sum(l\_extendedprice * (1 - l\_discount)) as revenue) from customer, 
orders, lineitem, supplier, nation, region where c\_custkey = o\_custkey and l\_orderkey = o\_orderkey and l\_suppkey = 
s\_suppkey and c\_nationkey = s\_nationkey and s\_nationkey = n\_nationkey and n\_regionkey = r\_regionkey and r\_name 
= 
``asia'' and o\_orderdate $>=$ ``1995-03-09'' and o\_orderdate $<$ ``1996-03-09'';

\noindent{\bf Q14}: \selectq{select} supp\_nation, cust\_nation, l\_shipdate, (sum (volume) as revenue) from (( 
select (n1.n\_name as supp\_nation), ( n2.n\_name as cust\_nation), l\_shipdate, ((l\_extendedprice * (1 - l\_discount)) 
as volume) from supplier, lineitem, orders, customer, (nation as n1), (nation as n2) where s\_suppkey = l\_suppkey and 
o\_orderkey = l\_orderkey and c\_custkey = o\_custkey and s\_nationkey = n1.n\_nationkey and c\_nationkey = 
n2.n\_nationkey and ( (n1.n\_name = ``india'' and n2.n\_name = ``united states'') or (n1.n\_name = ``united states'' 
and 
n2.n\_name = ``india'') ) and l\_shipdate $>$ `1995-01-01' and l\_shipdate $<$ `1996-12-31' ) as shipping);

\noindent{\bf Q15}: \selectq{select} (sum(l\_extendedprice * l\_discount) as revenue) from lineitem where l\_shipdate 
$>=$ ``1994-04-15'' and l\_shipdate $<$ ``1995-04-15'' and l\_discount $>$ 0.04 - 0.01 and l\_discount $<$ 0.04 + 0.01 
and 
l\_quantity $<$ 20 ;

\noindent{\bf Q16}: \selectq{select} o\_orderpriority, (count(*) as order\_count) from orders where o\_orderdate $>=$ 
``1995-03-09'' and o\_orderdate $<$ ``1995-06-09'';

\noindent{\bf Q17}: \selectq{select} l\_shipmode, o\_orderpriority from orders, lineitem where o\_orderkey = 
l\_orderkey and (l\_shipmode = ``ship'' or l\_shipmode = ``air'') and l\_commitdate $<$ l\_receiptdate and l\_shipdate 
$<$ 
l\_commitdate and l\_receiptdate $>=$ ``1995-01-01'' and l\_receiptdate $<$ ``1996-01-01'';

\noindent{\bf Q18}: \insertq{insert} into customer ( c\_custkey  , c\_name , c\_address , c\_nationkey , c\_phone , 
c\_acctbal ,c\_mktsegment, c\_comment)  values (91639738 , ``sumpil'', ``renol'', 93793619 , ``9242'', 234, 
``pilsum'', ``rolen'') , ( 91639737 , ``abc'', ``def'', 93793619 , ``1234'', 234, ``yhbdsra'', ``afgsdf''), 
(96244913, ``pkjhbc'', ``mnhgre'', 93793619 , ``9543'', 234, ``qaxcvf'', ``iomnbgf''), ( 96244914, ``yuthgbvfg'', 
``qgytrevd'', 93793619 , ``75345'', 234, ``liyhvdrt'', ``qfgkdyv''), (96244915, ``ramnabfubt'',\\ ``njhiyfcvh'', 
93793619 , ``126789'', 234, ``summinhsve'',\\ ``qgjorutbbs'');

\noindent{\bf Q19}: \updateq{update} customer set  c\_name=``ashdehhrbeki'' where c\_name = ``sjadfd'';

\noindent{\bf Q20}: \updateq{update} customer set c\_name = ``sjadfd'', c\_address = ``asndkwewhrwoer'', 
c\_phone = ``3245456458'', c\_acctbal=22 ,c\_mktsegment = ``ajshuejre'', c\_comment = ``asjhdeke'' where c\_nationkey = 
93793619;

\noindent{\bf Q21}: \deleteq{delete} from customer where c\_nationkey = 93793619;

\noindent{\bf Q22}: \insertq{insert}  into supplier  (s\_suppkey, s\_name, s\_address, s\_nationkey, s\_phone, 
s\_acctbal, s\_comment) select ( c\_custkey + 1000 ), c\_name, c\_address, c\_nationkey, c\_phone, c\_acctbal, 
c\_comment from customer where c\_nationkey = 93793619;

\noindent{\bf Q23}: \selectq{select} l\_orderkey, (sum ( l\_extendedprice) as revenue), o\_orderdate, o\_shippriority 
from customer, orders, lineitem where c\_mktsegment = ``automobile'' and c\_custkey = o\_custkey and l\_orderkey = 
o\_orderkey and o\_orderdate $<$ ``1995-03-09'' and l\_shipdate $>$ ``1995-03-09'';

\noindent{\bf Q24}: \selectq{select} (sum ( l\_extendedprice * (1 - l\_discount )) as promo\_revenue) from lineitem, 
part where l\_partkey = p\_partkey and l\_shipdate $>=$ ``1995-01-01'' and l\_shipdate $<$ ``1995-02-01'';

\noindent{\bf Q25}: \selectq{select} * from customer;

\noindent{\bf Q26}: \deleteq{delete} from supplier where s\_nationkey = 93793619;

\noindent{\bf Q27}: \selectq{select} p\_brand, p\_type, p\_size, (count ( ps\_suppkey ) as supplier\_cnt ) from 
partsupp, part where p\_partkey = ps\_partkey and p\_brand $<>$ `brand\#34' and p\_type not like `medium brushed brass' 
and ( p\_size = 22 or p\_size = 47 or p\_size = 30 or p\_size = 29 or p\_size = 11 or p\_size = 37 or p\_size = 42 or 
p\_size = 34 or p\_size = 40 );

\noindent{\bf Q28}: \selectq{select} * from part;

\noindent{\bf Q29}: \selectq{select} ps\_partkey, (sum(ps\_supplycost * ps\_availqty) as value) from partsupp, 
supplier, nation where ps\_suppkey = s\_suppkey and s\_nationkey = n\_nationkey and n\_name = ``india'';

\noindent{\bf Q30}: \selectq{select} c\_custkey, c\_name, (sum ( l\_extendedprice * (1 - l\_discount)) as revenue), 
c\_acctbal, n\_name, c\_address, c\_phone, c\_comment from customer, orders, lineitem, nation where c\_custkey = 
o\_custkey and l\_orderkey = o\_orderkey and o\_orderdate $>=$ ``1995-01-01'' and o\_orderdate $<$ ``1995-04-01'' and 
l\_returnflag = `r' and c\_nationkey = n\_nationkey;

\noindent{\bf Q31}: \updateq{update} supplier set s\_acctbal = (select sum(c\_custkey) from (customer as c), (nation 
as n) where c.c\_nationkey = n.n\_nationkey), s\_phone = ( select c\_phone from customer where c\_custkey= 91639738), 
s\_comment = ``asdherbejhrbeh''   where s\_suppkey = 100001;

\noindent{\bf Q32}: \selectq{select} s\_suppkey, n\_name, s\_name from supplier, nation where supplier.s\_nationkey = 
nation.n\_nationkey;

\noindent{\bf Q33}: \selectq{select} s\_suppkey, n\_name, s\_name from (( select * from supplier ) as sup ), nation 
where sup.s\_nationkey = nation.n\_nationkey;

\noindent{\bf Q34}: \selectq{select} s\_suppkey, s\_name, s\_address, s\_phone, total\_revenue from supplier, 
(( select ( l\_suppkey), (max ( l\_extendedprice * (1 - l\_discount)) as total\_revenue) from lineitem where 
l\_shipdate $>=$ ``1995-01-01'' and l\_shipdate $<$ ``1995-04-01'') as sup) where s\_suppkey = sup.l\_suppkey;

\noindent{\bf Q35}: \selectq{select} sn.s\_name, rn.r\_name from (( select n\_name, s\_name from (supplier as sup ), 
nation where sup.s\_nationkey = nation.n\_nationkey) as sn), (( select n\_name, r\_name from ( region as reg ), nation 
where reg.r\_regionkey = nation.n\_regionkey ) as rn ) where sn.n\_name = rn.n\_name;

\noindent{\bf Q36}: \selectq{select} * from partsupp;

\noindent{\bf Q37}: \selectq{select} * from orders;

\noindent{\bf Q38}: \selectq{select} nation, o\_year, (sum (amount) as sum\_profit) from (( select ( n\_name as 
nation), (o\_orderdate as o\_year), ((( l\_extendedprice * (1 - l\_discount )) - ( ps\_supplycost * l\_quantity )) as 
amount ) from part, supplier, lineitem, partsupp, orders, nation where s\_suppkey = l\_suppkey and ps\_suppkey = 
l\_suppkey and ps\_partkey = l\_partkey and p\_partkey = l\_partkey and o\_orderkey = l\_orderkey and s\_nationkey = 
n\_nationkey and p\_name like `\%ab\%' ) as profit);

\noindent{\bf Q39}: \selectq{select} l\_returnflag, l\_linestatus, (sum (l\_quantity) as sum\_qty), 
(sum (l\_extendedprice) as sum\_base\_price), (avg (l\_quantity) as avg\_qty), (avg (l\_extendedprice) as avg\_price), 
(avg (l\_discount) as avg\_disc) from lineitem where l\_quantity $<=$ 20;

\noindent{\bf Q40}: \selectq{select} c\_orders.c\_custkey, (count(*) as custdist) from (( select c\_custkey, 
o\_orderkey from customer, orders where c\_custkey = o\_custkey and o\_comment not like `\%fi\%al\%' ) as c\_orders); 
select * from customer;

\noindent{\bf Q41}: \deleteq{delete} from supplier where s\_suppkey $>$ 20000;

\noindent{\bf Q42}: \insertq{insert} into supplier ( s\_suppkey, s\_name, s\_address, s\_nationkey, s\_phone, 
s\_acctbal, s\_comment ) select ( c\_custkey + 100000 ), n\_name, c\_address, c\_nationkey, c\_phone, c\_acctbal, 
c\_comment from ( customer as c), ( nation as n ) where c.c\_nationkey = n.n\_nationkey ;

\noindent{\bf Q43}: \selectq{select} * from lineitem; 
\end{appendices}


\begin{thebibliography}{10}

\bibitem{privacy}
{By 2021, 75 percent of public blockchains will suffer privacy poisoning --
  inserted personal data that renders the blockchain noncompliant with privacy
  laws.}
\newblock
  \url{https://www.dig-in.com/list/gartners-top-10-tech-predictions-for-2019-and-beyond}.

\bibitem{restapi}
{Hyperledger Composer}.
\newblock \url{https://hyperledger.github.io/composer/v0.19/index.html}.

\bibitem{nyschool}
{NY school official accused of changing grades}.
\newblock
  \url{http://www.fox5ny.com/news/new-rochelle-grade-tampering-scandal}.

\bibitem{VormetricReport2015}
Trends and future directions in data security – 2015 vormetric insider threat
  report.
\newblock Technical report, 2015.
\newblock
  \url{https://dtr.thalesesecurity.com/insiderthreat/2015/pdf/2015-vormetric-insider-threat-press-deck-v3.pdf}.

\bibitem{haystax}
{2017 Insider Threat Study}, 2017.
\newblock
  \url{https://haystax.com/blog/whitepapers/insider-attacks-industry-survey/}.

\bibitem{iowa2017}
{Former University of Iowa student nabbed in high-tech cheating scheme}, 2017.
\newblock
  \url{http://www.nydailynews.com/news/national/student-arrested-stealing-tests-changing-grades-article-1.3595691}.

\bibitem{maryland2017}
{Probe finds late grade changes for 5,500 in Prince George’s}, 2017.
\newblock
  \url{https://www.washingtonpost.com/local/education/probe-finds-late-grade-changes-for-5500-in-prince-georges/2017/11/03/5e54
  e10c-be62-11e7-959c-fe2b598d8c00\_story.html}.

\bibitem{evm3}
{AAP Alleges Mass Deletion Of Votes, EVM Tampering At All-Party Meet}, 2018.
\newblock
  \url{https://www.ndtv.com/delhi-news/aap-alleges-mass-deletion-of-votes-evm-tampering-at-all-party-meet-1960579}.

\bibitem{bigchaindb18}
Bigchaindb 2.0: The blockchain database.
\newblock {\em white paper}, 2018.
\newblock
  \url{https://www.bigchaindb.com/whitepaper/bigchaindb-whitepaper.pdf}.

\bibitem{chaindb}
Chaindb: A peer-to-peer database system, 2018.
\newblock \url{https://bitpay.com/chaindb.pdf}.

\bibitem{evm2}
{Congress demands JPC probe over alleged irregularities in Telangana polls},
  2018.
\newblock
  \url{https://www.indiatoday.in/elections/story/congress-demands-jpc-probe-over-alleged-irregularities-in-telangana-polls-1409998-2018-12-15}.

\bibitem{chf}
Cryptographic hash functions, 2018.
\newblock \url{https://en.wikipedia.org/wiki/Cryptographic_hash_function}.

\bibitem{georgia2018}
{Former UGA student hacked into system to change grade}, 2018.
\newblock
  \url{http://www.fox5atlanta.com/news/police-former-uga-student-hacked-into-system-to-change-grade}.

\bibitem{fabric}
Hyperledger fabric, 2018.
\newblock \url{https://www.hyperledger.org/projects/fabric}.

\bibitem{catechreport}
{Insider Threat -- CA Technologies}, 2018.
\newblock
  \url{https://www.ca.com/content/dam/ca/us/files/ebook/insider-threat-report.pdf}.

\bibitem{ledgerdb}
{LedgerDB github repo.}, 2018.
\newblock \url{https://github.com/ledgerdb/ledgerdb}.

\bibitem{mongodb18}
{MongoDB -- Opensource Document Database}, 2018.
\newblock \url{https://www.mongodb.com/}.

\bibitem{evm}
{Opposition raises EVM tampering claims in Kairana, Noorpur}, 2018.
\newblock
  \url{https://economictimes.indiatimes.com/news/politics-and-nation/opposition-raises-evm-tampering-claims-in-kairana-noorpur/
  articleshow/64351367.cms}.

\bibitem{sqlite}
{SQLite Database}, 2018.
\newblock \url{https://www.sqlite.org/index.html}.

\bibitem{tendermint18}
Tendermint, 2018.
\newblock \url{https://tendermint.com/}.

\bibitem{tpch}
{TPC-H Benchmark}, 2018.
\newblock \url{http://www.tpc.org/tpch/}.

\bibitem{cyberDeception}
S.~Achleitner, T.~La~Porta, P.~McDaniel, S.~Sugrim, S.~V. Krishnamurthy, and
  R.~Chadha.
\newblock Cyber deception: Virtual networks to defend insider reconnaissance.
\newblock In {\em Proceedings of the 8th ACM CCS International Workshop on
  Managing Insider Security Threats}, MIST '16, pages 57--68, New York, NY,
  USA, 2016. ACM.

\bibitem{4547660}
G.~Ali, N.~A. Shaikh, and Z.~A. Shaikh.
\newblock Towards an automated multiagent system to monitor user activities
  against insider threat.
\newblock In {\em 2008 International Symposium on Biometrics and Security
  Technologies}, pages 1--5, April 2008.

\bibitem{4267567}
Q.~Althebyan and B.~Panda.
\newblock A knowledge-base model for insider threat prediction.
\newblock In {\em 2007 IEEE SMC Information Assurance and Security Workshop},
  pages 239--246, June 2007.

\bibitem{Baracaldo2012}
N.~Baracaldo and J.~Joshi.
\newblock {A Trust-and-risk Aware RBAC Framework: Tackling Insider Threat}.
\newblock In {\em Proceedings of the 17th ACM Symposium on Access Control
  Models and Technologies}, SACMAT '12, pages 167--176, New York, NY, USA,
  2012. ACM.

\bibitem{barbara02}
D.~Barbar{\'a}, R.~Goel, and S.~Jajodia.
\newblock {Mining malicious corruption of data with hidden Markov models}.
\newblock In {\em Research Directions in Data and Applications Security}, pages
  175--189. Springer, 2003.

\bibitem{bertino05}
E.~Bertino, E.~Terzi, A.~Kamra, and A.~Vakali.
\newblock {Intrusion detection in RBAC-administered databases}.
\newblock In {\em Computer security applications conference, 21st annual},
  pages 10--pp. IEEE, 2005.

\bibitem{Bishop2008}
M.~Bishop, S.~Engle, S.~Peisert, S.~Whalen, and C.~Gates.
\newblock We have met the enemy and he is us.
\newblock In {\em Proceedings of the 2008 New Security Paradigms Workshop},
  NSPW '08, pages 1--12, New York, NY, USA, 2008. ACM.

\bibitem{Chagarlamudi2009}
M.~Chagarlamudi, B.~Panda, and Y.~Hu.
\newblock Insider threat in database systems: Preventing malicious users'
  activities in databases.
\newblock In {\em 2009 Sixth International Conference on Information
  Technology: New Generations}, pages 1616--1620, April 2009.

\bibitem{demids}
C.~Y. Chung, M.~Gertz, and K.~Levitt.
\newblock {Demids: A misuse detection system for database systems}.
\newblock In {\em Integrity and Internal Control in Information Systems}, pages
  159--178. Springer, 2000.

\bibitem{CollinsCommonSense2016}
M.~Collins, M.~Theis, R.~Trzeciak, J.~Strozer, J.~Clark, D.~Costa, T.~Cassidy,
  M.~Albrethsen, and A.~Moore.
\newblock Common sense guide to mitigating insider threats.
\newblock Technical Report CMU/SEI-2016-TR-015, Software Engineering Institute,
  Carnegie Mellon University, Pittsburgh, PA, 2016.

\bibitem{Crampton2010TowardsAA}
J.~Crampton and M.~Huth.
\newblock Towards an access-control framework for countering insider threats.
\newblock In {\em Insider Threats in Cyber Security}, 2010.

\bibitem{composerpoorperf}
G.~Engstrand.
\newblock {Evaluating Hyperledger Composer}.
\newblock \url{https://www.infoq.com/articles/evaluating-hyperledger-composer}.

\bibitem{Garfinkel2002}
R.~Garfinkel, R.~Gopal, and P.~Goes.
\newblock Privacy protection of binary confidential data against deterministic,
  stochastic, and insider threat.
\newblock {\em Management Science}, 48(6):749--764, 6 2002.

\bibitem{Gopal2006}
R.~Garfinkel, R.~Gopal, and D.~Rice.
\newblock New approaches to disclosure limitation while answering queries to a
  database: Protecting numerical confidential data against insider threat based
  on data or algorithms.
\newblock In {\em Proceedings of the 39th Annual Hawaii International
  Conference on System Sciences (HICSS'06)}, volume~6, pages 125a--125a, Jan
  2006.

\bibitem{GopalGG02}
R.~D. Gopal, R.~S. Garfinkel, and P.~B. G{\'{o}}es.
\newblock Confidentiality via camouflage: The {CVC} approach to disclosure
  limitation when answering queries to databases.
\newblock {\em Operations Research}, 50(3):501--516, 2002.

\bibitem{Handschuh2005}
H.~Handschuh.
\newblock {\em SHA Family (Secure Hash Algorithm)}, pages 565--567.
\newblock Springer US, Boston, MA, 2005.

\bibitem{hashemi08}
S.~Hashemi, Y.~Yang, D.~Zabihzadeh, and M.~Kangavari.
\newblock Detecting intrusion transactions in databases using data item
  dependencies and anomaly analysis.
\newblock {\em Expert Systems}, 25(5):460--473, 2008.

\bibitem{EthernityDB}
S.~Helmer, M.~Roggia, N.~E. Ioini, and C.~Pahl.
\newblock {EthernityDB -- Integrating Database Functionality into a
  Blockchain}.
\newblock In {\em {New Trends in Databases and Information Systems}}, 2018.

\bibitem{insider18}
I.~Homoliak, F.~Toffalini, J.~Guarnizo, Y.~Elovici, and M.~Ochoa.
\newblock Insight into insiders: {A} survey of insider threat taxonomies,
  analysis, modeling, and countermeasures.
\newblock {\em CoRR}, abs/1805.01612, 2018.

\bibitem{hu04}
Y.~Hu and B.~Panda.
\newblock A data mining approach for database intrusion detection.
\newblock In {\em Proceedings of the 2004 ACM symposium on Applied computing},
  pages 711--716. ACM, 2004.

\bibitem{GusJabbour2010StoppingTI}
G.~G. Jabbour and D.~A. Menasc{\'e}.
\newblock Stopping the insider threat : the case for implementing integrated
  autonomic defense mechanisms in computing systems.
\newblock 2010.

\bibitem{Kandias10}
M.~Kandias, A.~Mylonas, N.~Virvilis, M.~Theoharidou, and D.~Gritzalis.
\newblock An insider threat prediction model.
\newblock In S.~Katsikas, J.~Lopez, and M.~Soriano, editors, {\em Trust,
  Privacy and Security in Digital Business}, pages 26--37, Berlin, Heidelberg,
  2010. Springer Berlin Heidelberg.

\bibitem{tendermint14}
J.~Kwon.
\newblock Tendermint: Consensus without mining.
\newblock {\em Draft v. 0.6, fall}, 2014.

\bibitem{lee00}
V.~C. Lee, J.~A. Stankovic, and S.~H. Son.
\newblock Intrusion detection in real-time database systems via time
  signatures.
\newblock In {\em Real-Time Technology and Applications Symposium, 2000. RTAS
  2000. Proceedings. Sixth IEEE}, pages 124--133. IEEE, 2000.

\bibitem{nakamoto}
S.~Nakamoto.
\newblock {Bitcoin: A Peer-to-Peer Electronic Cash System}.
\newblock 2008.
\newblock \url{https://bitcoin.org/bitcoin.pdf}.

\bibitem{Panigrahi2013}
S.~Panigrahi, S.~Sural, and A.~K. Majumdar.
\newblock Two-stage database intrusion detection by combining multiple evidence
  and belief update.
\newblock {\em Information Systems Frontiers}, 15(1):35--53, Mar 2013.

\bibitem{Pramanik2004}
S.~Pramanik, V.~Sankaranarayanan, and S.~Upadhyaya.
\newblock Security policies to mitigate insider threat in the document control
  domain.
\newblock In {\em 20th Annual Computer Security Applications Conference}, pages
  304--313, Dec 2004.

\bibitem{dbbook}
R.~Ramakrishnan and J.~Gehrke.
\newblock {\em Database Management Systems}.
\newblock McGraw-Hill, Inc., New York, NY, USA, 3 edition, 2003.

\bibitem{4024507}
V.~Sankaranarayanan, S.~Pramanik, and S.~Upadhyaya.
\newblock Detecting masquerading users in a document management system.
\newblock In {\em 2006 IEEE International Conference on Communications},
  volume~5, pages 2296--2301, June 2006.

\bibitem{srivastava06}
A.~Srivastava, S.~Sural, and A.~K. Majumdar.
\newblock {Weighted intra-transactional rule mining for database intrusion
  detection}.
\newblock In {\em Pacific-Asia Conference on Knowledge Discovery and Data
  Mining}, pages 611--620. Springer, 2006.

\bibitem{veritytechrep}
S.~S. Srivastava, M.~Atre, S.~Sharma, R.~Gupta, and S.~K. Shukla.
\newblock {Verity Technical Report -- Additional Experimental Results}.
\newblock Technical report.
\newblock
  \url{https://www.cs.ox.ac.uk/people/medha.atre/papers/veritytechrep.pdf}.

\bibitem{Vance:2015:IAT:2871313.2871318}
A.~Vance, P.~B. Lowry, and D.~Eggett.
\newblock Increasing accountability through user-interface design artifacts: A
  new approach to addressing the problem of access-policy violations.
\newblock {\em MIS Q.}, 39(2):345--366, June 2015.

\bibitem{Wu2009}
G.~Z. Wu, S.~L. Osborn, and X.~Jin.
\newblock Database intrusion detection using role profiling with role
  hierarchy.
\newblock In W.~Jonker and M.~Petkovi{\'{c}}, editors, {\em Secure Data
  Management}, pages 33--48, Berlin, Heidelberg, 2009. Springer Berlin
  Heidelberg.

\bibitem{Wu2011}
J.~Wu, J.~Zhou, J.~Ma, S.~Mei, and J.~Ren.
\newblock An active data leakage prevention model for insider threat.
\newblock In {\em Intelligence Information Processing and Trusted Computing
  (IPTC), 2011 2nd International Symposium on}, pages 39--42. IEEE, 2011.

\end{thebibliography}
\end{document}